%% file: recurr_arxiv.tex
\documentclass[pra
	       ,nofootinbib
	       ,floatfix
	       ,superscriptaddress
	       ,twocolumn
	       ]{revtex4-1}
	       
\usepackage{times}

\usepackage{amsmath} 
\usepackage{graphicx} 
\usepackage{color} 
\usepackage[utf8]{inputenc}
\usepackage{siunitx}
\usepackage[hidelinks]{hyperref}
\usepackage[hidelinks]{hyperref}

\newcommand*{\angfreq} [2] {2\pi \cdot \SI{#1}{#2\hertz}}

\newcommand{\trec}[0]{\ensuremath{t_\mathrm{rec}}}
\newcommand{\Hrec}[0]{\ensuremath{H_\mathrm{rec}}}
\newcommand{\Teff}[0]{\ensuremath{T_\mathrm{eff}}}

\newcommand{\cohfact}[0]{\ensuremath{\langle \mathrm{cos}(\varphi)\rangle}}
\newcommand{\Cnorm}{\ensuremath{\mathcal{C}^2/\langle \mathcal{C}^2 \rangle}}
\newcommand{\Cbase}{\ensuremath{C_\mathrm{base}}}
\newcommand{\zc}{\ensuremath{\bar{z}_c}}

\newcommand{\SOM}{SM}

\newcommand{\fig}[1]{fig.~\ref{fig:#1}}
\newcommand{\Fig}[1]{Fig.~\ref{fig:#1}}

\newcommand{\SOMfirstPICposition}{\begin{figure}[b]}

\begin{document}

\title{Recurrences in an isolated quantum many-body system}

\author{Bernhard~Rauer}
\affiliation{Vienna Center for Quantum Science and Technology, Atominstitut, TU Wien, Stadionallee 2, 1020 Vienna, Austria}

\author{Sebastian~Erne}
\affiliation{Vienna Center for Quantum Science and Technology, Atominstitut, TU Wien, Stadionallee 2, 1020 Vienna, Austria}
\affiliation{Institut f\"{u}r Theoretische Physik, Universit\"{a}t Heidelberg, Philosophenweg 16, 69120 Heidelberg, Germany}
\affiliation{Kirchhoff-Institut f\"{u}r Physik, Universit\"{a}t Heidelberg, Im Neuenheimer Feld 227, 69120 Heidelberg, Germany}

\author{Thomas~Schweigler} 
\affiliation{Vienna Center for Quantum Science and Technology, Atominstitut, TU Wien, Stadionallee 2, 1020 Vienna, Austria}

\author{Federica~Cataldini}
\affiliation{Vienna Center for Quantum Science and Technology, Atominstitut, TU Wien, Stadionallee 2, 1020 Vienna, Austria}

\author{Mohammadamin~Tajik}
\affiliation{Vienna Center for Quantum Science and Technology, Atominstitut, TU Wien, Stadionallee 2, 1020 Vienna, Austria}
 
    
\author{J\"org~Schmiedmayer}
\email[]{schmiedmayer@atomchip.org}
\affiliation{Vienna Center for Quantum Science and Technology, Atominstitut, TU Wien, Stadionallee 2, 1020 Vienna, Austria}

\date{\today}

\begin{abstract}
\input{abstract}
\end{abstract}

\maketitle

\input{recurr_text}

\vspace{5mm}
\noindent{\bf Acknowledgments:} \\
\input{acknowledgements}

\bibliography{biblio_final}

\clearpage
\onecolumngrid

\renewcommand{\thefigure}{S\arabic{figure}}
\setcounter{figure}{0}

\begin{center}
  \LARGE
  \textbf{Supplementary Materials} 
\end{center}



\input{som_sec1_exp_basics.tex}
\input{som_sec2_data_analysis.tex}
\input{som_sec3_theory.tex}
\input{som_sec4_numerics.tex}


\end{document}

%% file: abstract.tex
Even though the evolution of an isolated quantum system is unitary, the complexity of interacting many-body systems prevents the observation of recurrences of quantum states for all but the smallest systems. For large systems one can not access the full complexity of the quantum states and the requirements to observe a recurrence in experiments reduces to being close to the initial state with respect to the employed observable. Selecting an observable connected to the collective excitations in one-dimensional superfluids, we demonstrate recurrences of coherence and long range order in an interacting quantum many-body system containing thousands of particles. This opens up a new window into the dynamics of large quantum systems even after they reached a transient thermal-like state.   




%% file: recurr_text.tex
The expectation that a non-equilibrium system evolves towards thermal equilibrium is deeply rooted in our daily experience and formed the beginning of statistical mechanics~\cite{BoltzmannH}. On the other hand, as formulated by Poincar\'e and Zermelo, a {\em finite} isolated physical system will recur  arbitrarily close to its initial state after a sufficiently long but finite time~\cite{Poincare1890,Zermelo1896}. The reconciliation of these seemingly contradicting statements is at the heart of the emergence of irreversible processes from reversible microscopic mechanics~\cite{terHaar1955}.


The above discussion can be transferred to the quantum domain. In analogy to Boltzmann's conjecture, already in 1929 von Neumann formulated a quasi-ergodic theorem for the evolution of the wave function~\cite{Neumann1929} and the equilibration of isolated quantum systems grew into an active field of research~\cite{Gogolin2016}. However, also in quantum mechanics a general recurrence theorem can be proven~\cite{Bocchieri1957,Percival1961}, explicitly showing that the wave function returns arbitrary close to its initial state. 


A beautiful example of recurrences in a simple quantum system is the prediction~\cite{Cummings1965,Eberly1980} and observation~\cite{Rempe1987} of collapse and revivals in the Jaynes-Cummings model of a single atom interacting with a coherent radiation field.
In interacting few-body systems collapse and revivals were observed for small samples of a few atoms trapped in optical lattices~\cite{Greiner2002a,Will2010}. 
For larger systems however, the complexity of the spectrum of eigenstates leads to exceedingly long recurrence times, in general prohibiting their observation. 

Moreover, for many-body systems it becomes exponentially difficult to observe the eigenstates directly.  Instead, one investigates these systems through much simpler measurements of (local) few-body observables $\mathcal{O}$. 
This poses the question if and under which condition the observation of recurrences becomes feasible. 
The system does not have to come back close to the exact initial configuration of many-body states, but only needs to give the same measurement results under the evaluation of $\mathcal{O}$. 

In quantum many-body systems such observables $\mathcal{O}$ can be chosen to reflect the collective excitations of the underlying quantum field theory description~\cite{Schweigler2017}. This dramatically reduces the complexity of the problem from a large number of constituents to a much smaller number of populated modes.
Designing the whole system such that the collective excitations can be described by a few commensurate modes, the observation of recurrences becomes feasible even for many-body systems containing thousands of interacting particles. 

Ultra cold gases~\cite{Proukakis2017} are an ideal starting point to study these fundamental phenomena at the border between quantum physics and statistical mechanics as they can be well isolated from the environment and excellent tools are available to prepare, manipulate and probe them.

As a model system we study coherence in one-dimensional (1D) superfluids. In first approximation, the corresponding many-body physics can be mapped to an effective low-energy description: a bosonic Luttinger liquid~\cite{Tomonaga1950a, Luttinger1963, Mattis1965, Giamarchi2004} where the collective excitations are free phonons. These phonons are directly related to phase fluctuations observed in interference of two 1D superfluids~\cite{Schweigler2017}. 

Loss of coherence and long range order proceeds through dephasing of these collective phononic excitations~\cite{Bistritzer2007,Kitagawa2011,Gring2012}. The dephased state emerges in a light-cone-like fashion~\cite{Langen2013} and is described by a generalized Gibbs ensemble~\cite{Langen2015a}. The long time behaviour depends on the spectrum of the collective phononic modes.
In an harmonic longitudinal confinement with trap frequency $\omega_z$, these phonon frequencies are non-commensurate $\omega_j = \omega_z \sqrt{j(j+1)/2}$~\cite{Petrov2000}, with $j$ being the mode index.  If the atoms are confined to a box shaped trap the phonon frequencies become commensurate and recurrences should be observable at short times~\cite{Geiger2014}. 

\begin{figure}[t]
  \centering
  \includegraphics[width=0.45\textwidth]{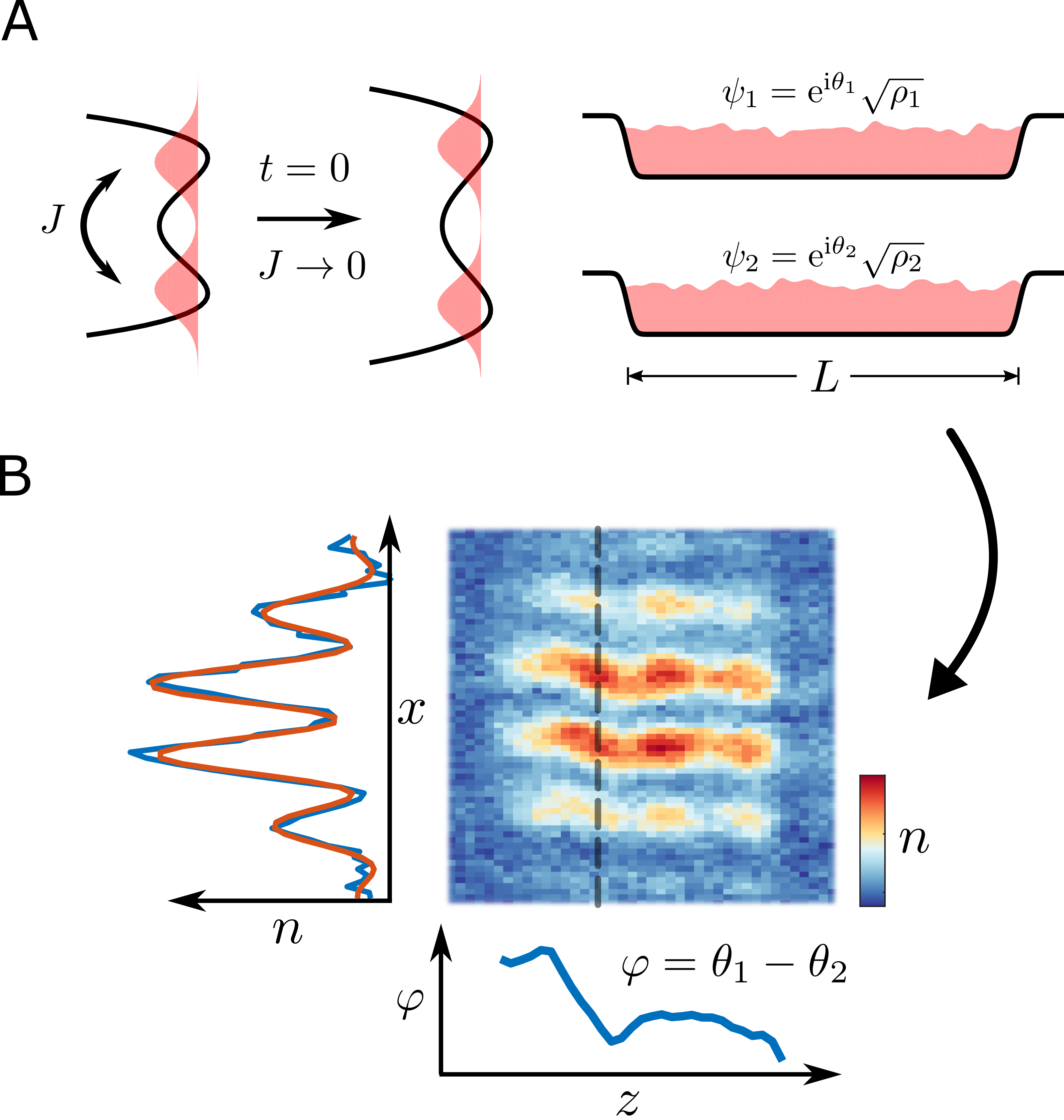}
  \caption{{\bf Schematics of the experiment and the measurement process.} ({\bf A}) Two coupled 1D superfluids $\psi_{1(2)}$ in a double well potential are taken out of equilibrium by a sudden quench of the tunnel coupling $J$ to zero. Along the longitudinal direction $z$ the system is confined to a box shaped potential of variable length $L$. ({\bf B}) The evolution of the decoupled system is measured through matter-wave interference in time-of-flight. A typical interference picture showing the atomic density $n$	is given. A sinusodal function is fitted to the local position of interference fringes for each position $z$. On the left we show an example of such a fit (red) for one slice (blue) indicated by the dashed line in the image. This gives access to the spatially resolved relative phase $\varphi(z) = \theta_1(z)-\theta_2(z)$ between the two superfluids (bottom).
}
  \label{fig:fig_1}
\end{figure}

We implement a box like confinement in an atom chip double well setup~\cite{Schumm2005} by adding hard walls to the very weak longitudinal harmonic confinement with the help of a blue detuned optical dipole potential (for details see \SOM{}).  
To observe recurrent dynamics we prepare a thermal equilibrium state in a coupled double well potential~\cite{Betz2011}. 
Typical samples have a linear density of about 70 atoms per \SI{}{\micro\meter} and depending on the box size between 2300 and 4800 atoms in each well, resulting in an interaction energy close to \SI{1}{\kilo\hertz}. 
The tunnel coupling $J$ is tuned to a regime where the phase between the two superfluids locks. This creates a system with a strongly correlated relative phase field $\varphi$ ($\cohfact \approx 1$). To initiate the non-equilibrium dynamics, the coupling is rapidly ramped to zero leaving the two gases to evolve independently (\fig{fig_1}). 


We observe the subsequent dynamics by matter-wave interferometry~\cite{Schumm2005} which gives direct access to the spatially resolved relative phase $\varphi(z)$ between the two superfluids. To study the coherence and long range order we evaluate the two-point correlation function~\cite{Langen2013} 
\begin{equation}
	\label{Eq_CorrFun}
	C(\bar z = z - z') = \langle  \mathrm{cos}(\varphi(z)-\varphi(z')) \rangle
\end{equation}

\noindent with the expectation value taken over many experimental realizations. 

\begin{figure}[t]
  \centering
  \includegraphics[width=0.42\textwidth]{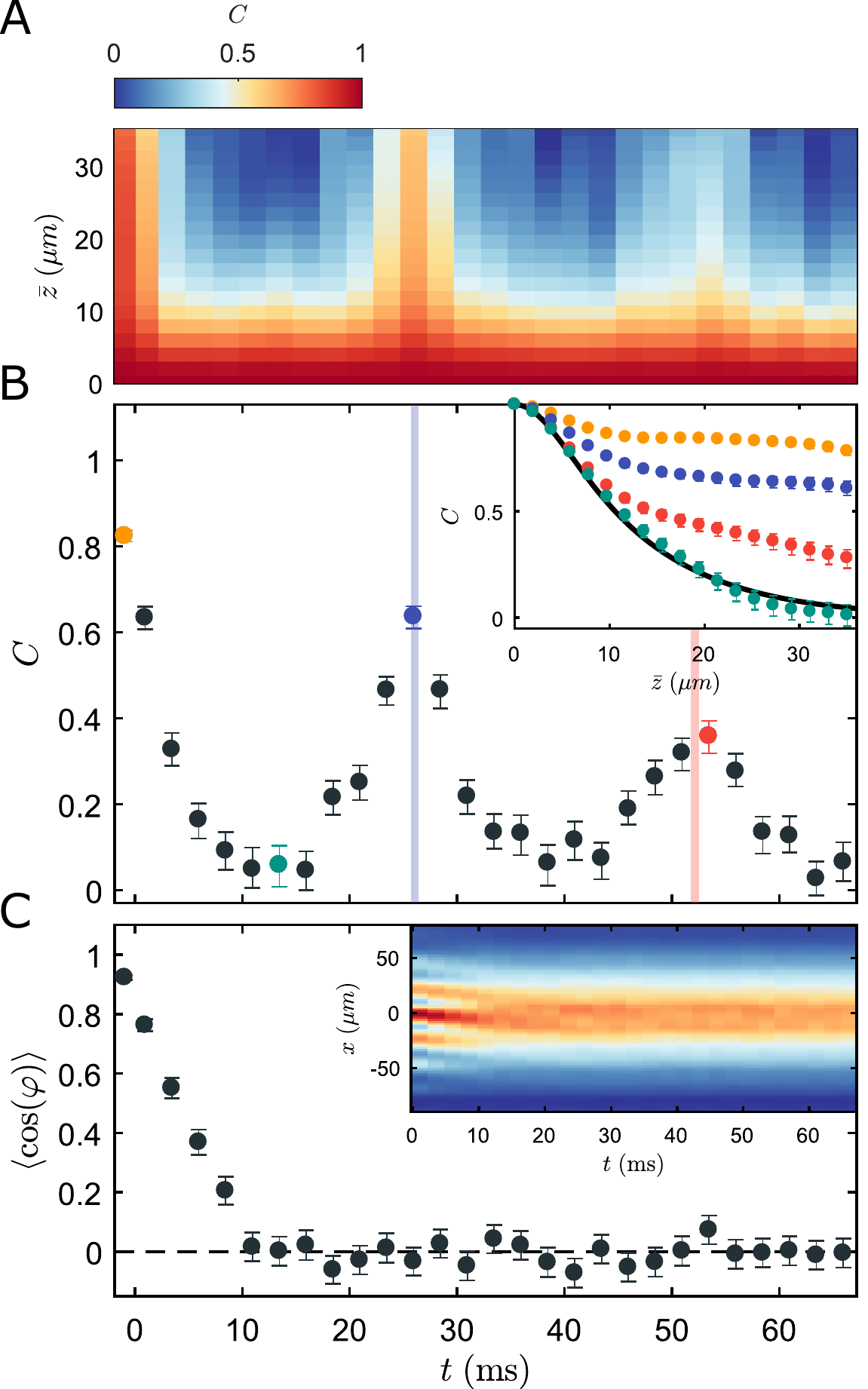}
  \caption{{\bf Dynamics after decoupling.}  ({\bf A}) Temporal evolution of the phase correlation function $C(\bar{z},t)$ after decoupling in a \SI{49}{\micro\meter} long box potential. ({\bf B}) A cut at $\zc = \SI{27.3}{\micro\meter}$ with the error bars giving the \SI{68}{\percent} confidence interval obtained from a bootstrap. The first and second recurrence are clearly visible and occur at the expected times $\trec = L/c$ and $2\,\trec$, as indicated by the vertical blue and red lines respectively. The inset shows the phase correlation function at the first (blue) and second (red) recurrence and compares them to the correlations in between (green) and the correlations in the initial state (orange). Corresponding points in (B) are colored accordingly. A thermal fit to the correlations in between the recurrences is given by the solid black line. ({\bf C}) Evolution of the coherence factor $\cohfact$ showing no sign of a recurrence. This illustrates that in contrast to the phase correlation function the averaged interference picture shows no recurrent behavior as shown in the inset (integrated along the $z$-axis).
	}
  \label{fig:fig_2}
\end{figure}

A typical temporal evolution of the phase correlations in a box trap of length $L=\SI{49}{\micro\meter}$ is shown in \fig{fig_2}A. Before the quench at $t=0$ the relative phase between the superfluids is locked and correlations are close to unity over the whole sample. Right after decoupling this long range order decays, reflecting the dephasing of the collective excitations. At the first minimum in \fig{fig_2}B the initial state is completely dephased and the system is indistinguishable from a thermal state (see inset). For a system with incommensurately spaced modes this dephased state persists for a long time, showing the emergence of statistical properties from the unitary quantum evolution~\cite{Gring2012}. In our system however, mode frequencies are designed to be commensurate and two partial recurrences of phase coherence are clearly visible in the subsequent evolution.


\begin{figure}[t]
  \centering
  \includegraphics[width=0.45\textwidth]{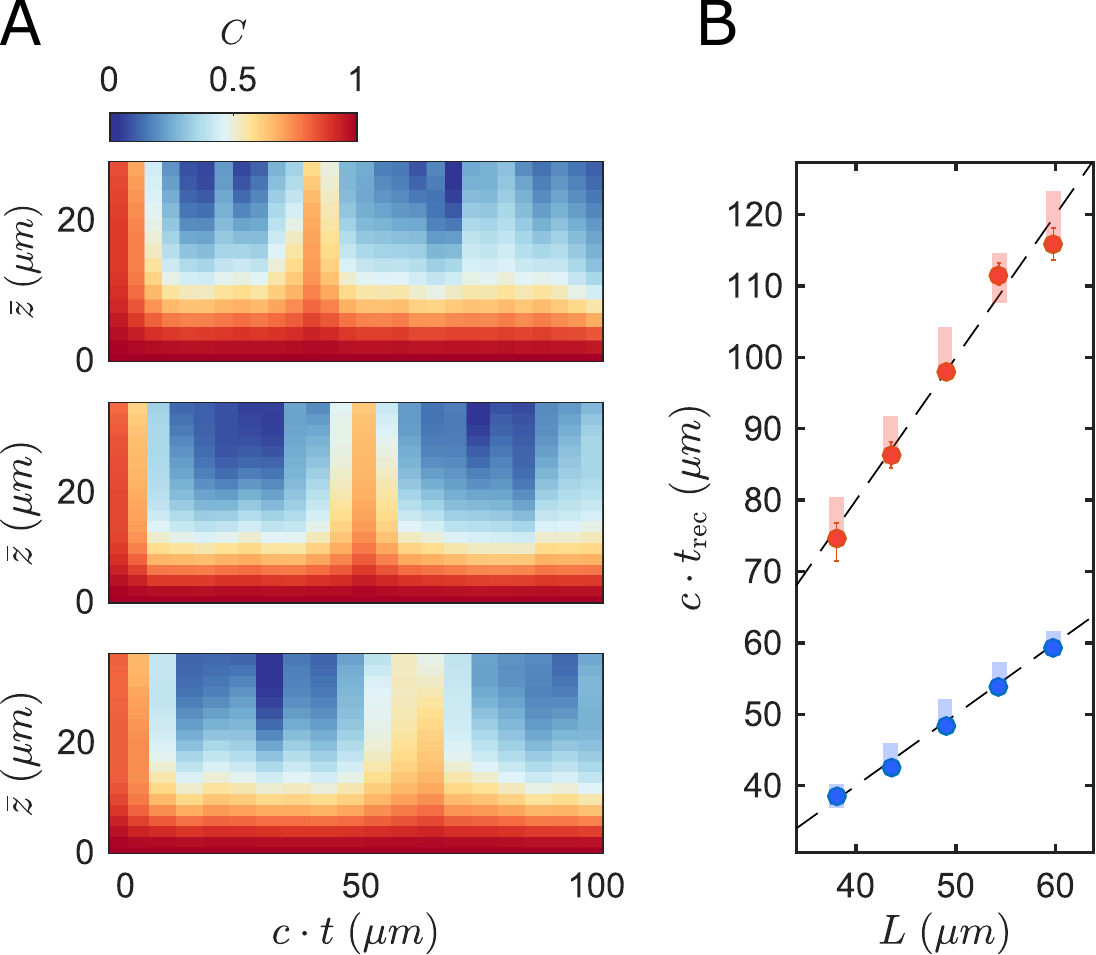}
  \caption{{\bf Comparison of the recurrences for different box lengths.} ({\bf A}) Phase correlations for three different box lengths $L =\;$\SIlist[list-units = single,list-final-separator = {, \text{and} }]{38;49;60}{\micro\meter} (top to bottom). The time axis is rescaled with the theoretical prediction for the speed of sound $c$. ({\bf B}) Recurrence time over the box length extracted from the phase correlations at $\zc = \SI{27.3}{\micro\meter}$ (see \SOM{}). For each box length the time of the first (blue) and second (red) recurrence is plotted and compared to the ideal linear scaling (dashed lines). The error bars give the \SI{68}{\percent} confidence interval obtained from a bootstrap while the shaded bars indicate the predictions of the Luttinger liquid model for our particular trap. The vertical extension of the bars corresponds to the uncertainty of the decoupling time while the horizontal extension is chosen arbitrarily.}
  \label{fig:fig_3}
\end{figure}

To understand the recurrence time we need to look at the dispersion relation of excitations. For a perfect hard-wall box confinement $\omega_j = \frac{c\pi}{L}\,j$, where $c$ is the speed of sound and $j$ is the mode index (see \SOM{}). More than commensurate, these modes are  equally spaced facilitating a recurrence at the earliest possible time $\frac{2\pi}{\Delta\omega} = \frac{2L}{c}$ with $\Delta\omega$ being the energy spacing between the modes. At this time the lowest lying mode finished a full rotation while the higher energy modes all performed an integer number of turns bringing all excitations back to their initial configuration. Half way to this full recurrence the system rephases to the mirrored initial state. As we initially start from a nearly flat relative phase profile and our observable $C$ is insensitive to the transformation $\varphi(z) \rightarrow \varphi(-z)$ this point is equivalent to the full recurrence. Therefore, the expected recurrence time for the correlations is given by $\trec = \frac{L}{c}$. In \fig{fig_2}B these times are indicated and agree well with the observed peaks in coherence.

It is interesting to note that the observation of recurrences depends on the choice of observables. The time evolution of the coherence factor $\cohfact$ shows no recurrence, as presented in \fig{fig_2}C. It relaxes during the initial dephasing dynamics and stays close to zero from then on. This is equivalent to observing that an averaged interference picture shows no revival of high contrast fringes as shown in the inset of \fig{fig_2}C. 
The reason behind this is a small random atom number imbalance between the two wells.
This imbalance originates from imperfections in the experiment, thermal fluctuations in the initial state, and the quantum noise of the decoupling process relevant for lower temperatures. 
It leads to an evolution of the  global relative phase that is different for each realization and therefore to vanishing interference contrast in the ensemble average. In contrast, the phase correlations (Eq.~\ref{Eq_CorrFun}) are insensitive to a global offset of the phase $\varphi(z)$, and recurrences can be observed.
This illustrates that a system can look perfectly equilibrated in one observable while exhibiting recurrent behavior in another.
  
To confirm the scaling of the recurrence time with the size of the system we vary the length of the box potential by changing the position of the dipole trap walls.  As shown in \fig{fig_3}A the recurrence is shifted to later times when the system size is increased. For this comparison the time axis was rescaled by the theoretical prediction for the speed of sound to make measurements with slightly different atomic densities comparable. Extracting the exact times of the first and second recurrence by fitting the peaks in the data (see \SOM{}) reveals the linear scaling with $L$ (see \fig{fig_3}B).


\begin{figure}[t]
  \centering
  \includegraphics[width=0.45\textwidth]{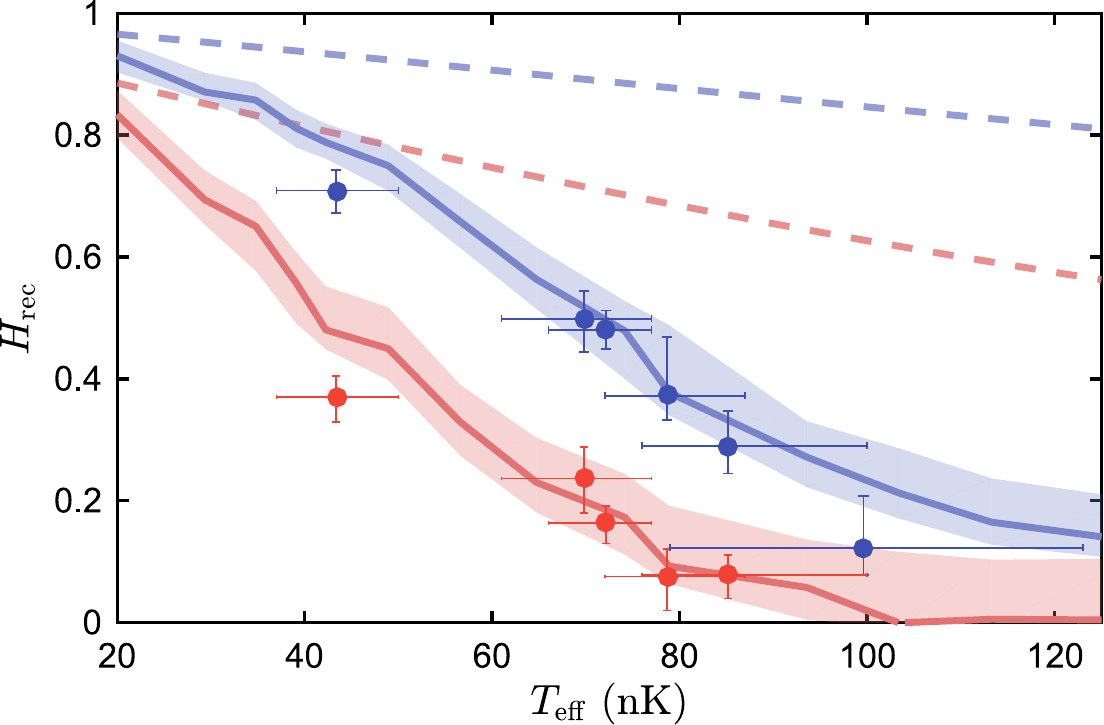}
  \caption{{\bf Temperature dependence of the recurrence height.} The data points represent measurements of the height $\Hrec$ of the first (blue) and second (red) recurrence in a box of length $L = \SI{49}{\micro\meter}$ for different effective temperatures $\Teff$ of the relative degrees of freedom. The height is extracted from fitting the peaks of the correlation function at $\zc = \SI{27.3}{\micro\meter}$ and the error bars give the \SI{68}{\percent} confidence interval obtained form a bootstrap. A fit of the full distribution function of measured interference contrasts at $t = \trec/2$ gives the effective temperature. The solid lines are results of GPE simulations analyzed in the same way as the experiment. The shaded area indicates the uncertainty due to the limited experimental statistics ($1\sigma$ deviation). The dashed lines are the predictions of the Luttinger liquid model for the experimental setup (see \SOM{} for details on the analysis). 
  }
  \label{fig:fig_4}
\end{figure}

Although the first two recurrences are clearly visible for all these measurements they are rapidly damped (see \fig{fig_2}A,B) and observing a third recurrence becomes unfeasible in most cases. To probe the decay of the recurrences in more detail, we studied the evolution for different initial temperatures in the $L = \SI{49}{\micro\meter}$ trap. Increasing the initial temperature the height of the observed recurrence with respect to the correlations of the initial state decreases rapidly as seen in \fig{fig_4}. The temperatures for this analysis are extracted from the full distribution functions of interference contrasts \cite{Hofferberth2008} for the completely dephased state in between the recurrences (see \SOM{}). 

To understand the origin of this damping with temperature we considered different theoretical descriptions of our system. 
We first investigated the low-energy effective description by solving the Luttinger liquid Hamiltonian.
This model describes the free propagation of phononic excitations on top of a stationary background density. For a homogenous background it would give perfect recurrence of phase coherence. For the comparison to the experimental data we consider an inhomogenous background density that reflects our box-like potential. In addition, we took into account the typical spread in particle number and imbalance between the wells, measured independently for the respective samples. While the recurrence times are well described by this model as seen in \fig{fig_3}B the damping, mainly coming from the shot-to-shot variation of the speed of sound in the experimental sample, is too weak to describe the experimental observations even for the lowest temperatures. 



As a second model we numerically simulated the dynamics using the Gross-Pitaevskii equation for finite temperature initial states (see \SOM{}). This description goes beyond the free phononic excitations of the Luttinger liquid and takes interactions between these quasi-particles into account. It agrees well with our experimental findings as seen in \fig{fig_4}. This shows that physics beyond the low-energy description becomes relevant and indicates that phonon-phonon interactions are responsible for the observed damping.  As the temperature is increased these processes get more important as higher energy modes are populated.


This illustrates that the observation of recurrences opens up a new perspective for non-equilibrium dynamics of quantum many-body systems~\cite{Cardy2014}. 
They are sensitive to the phase relations between the involved modes, and give different insight into relaxation processes compare to the standard probe of mode occupations. In our model case, the decay of recurrences with temperature clearly shows that physics beyond Luttinger liquid sets in quite quickly. A combined study looking at recurrences and mode occupations will shed light into the fundamental quantum processes in the relaxation of 1D systems.

%% file: acknowledgements.tex
We acknowledge discussions with I. Mazets, J. Berges, T. Gasenzer. This work was supported by the EU through the EU-FET Proactive grant AQuS, Project No. 640800, and the ERC advanced grant QuantumRelax, and by the Austrian Science Fund (FWF) through the doctoral programme CoQuS (\textit{W1210}) (B.R., T.S., F.C.). This work is supported by the SFB 1225 `ISOQUANT' financed by the German Research Foundation (DFG) and the FWF.


%% file: som_sec1_exp_basics.tex
\section{Experimental details}

\subsection{Setup}
A standard combination of laser cooling, evaporative cooling, and magnetic trapping is used to bring samples of neutral $^{87}$Rb atoms to degeneracy. The final confinement on an atom chip~\cite{Folman2000,Reichel2011} is created by a combination of static and radio-frequency (RF) magnetic fields, tuned to realize a dressed-state double well (DW) potential~\cite{Hofferberth2006,Lesanovsky2006a}. The trap consists of two parallel highly elongated harmonic wells each with a trap frequency of $\omega_\perp \simeq \angfreq{1.4}{\kilo}$ in the two tightly confined directions and $\omega_z \simeq \angfreq{7}{}$ in the elongated direction. By tuning the amplitude of the oscillating field the barrier height between the wells can be changed, controlling the tunnel-coupling $J$ between them.

\subsection{Box confinement}
A box shaped confinement along the elongated direction of the wells is created by superposing the harmonic magnetic trap with a blue detuned (\SI{767}{\nano\metre}) dipole trap. For this purpose we shine a Gaussian beam onto an exchangeable wire mask blocking a stripe of variable width from the intensity in the central part of the beam. This mask is imaged onto the atom clouds from a direction perpendicular to the weakly confined axis of the system. A schematic of the setup is depicted in \fig{box_potential}A. The resulting potential adds two steep walls to the shallow harmonic magnetic confinement $\omega_z$. At the walls the optical potential rises in \SI{3.1}{\micro\meter} from \SI{10}{\percent} to \SI{90}{\percent} of its total height, which is about $\SI{1.3}{\kilo\hertz}$ .

\Fig{box_potential}B shows a typical measured longitudinal density profile of a single superfluid in the box trap. Taking the resolution of the imaging system into account the measured steepness of the atom distribution at the edges of the profile is consistent with a diffraction limited projection of the shadow mask.

\SOMfirstPICposition
  \centering
  \includegraphics[width=0.95\textwidth]{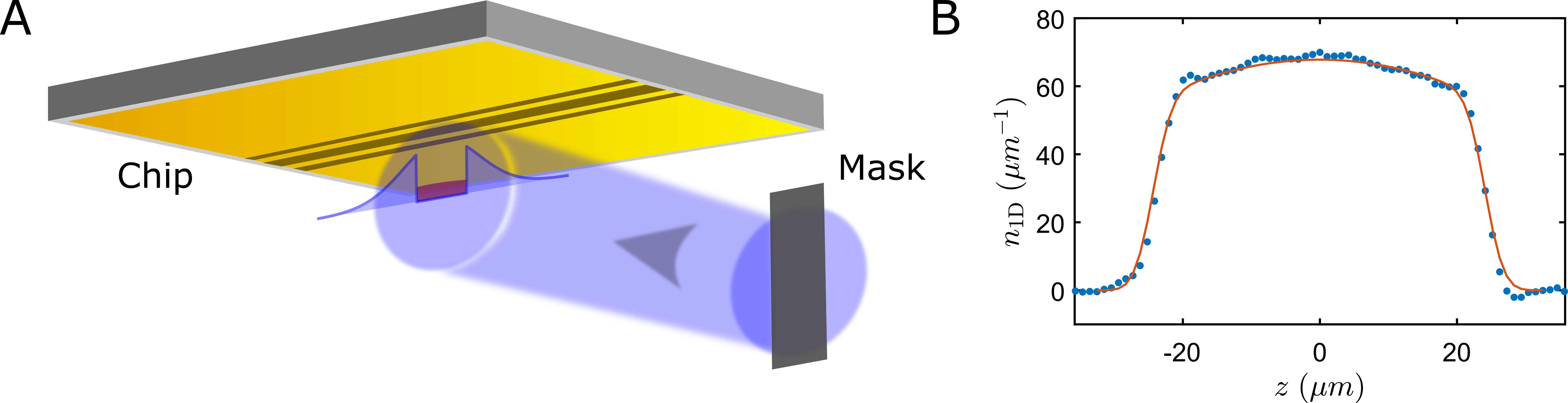}
  \caption{({\bf A}) Schematics of the setup shaping the box potential. A Gaussian beam of blue detuned laser light (blue) is cut in two by a mask realizing two steep potential walls in the plane of the atoms (red). ({\bf B}) Atomic density in the \SI{49}{\micro\meter} wide box potential after \SI{2}{\milli\second} time-of-flight expansion. An expected density profile obtained from a 1D Gross-Pitaevskii equation at $T=0$ with the imaging resolution taken into account is shown in red.
}
  \label{fig:box_potential}
\end{figure}

\subsection{Experimental procedure}
To prepare the initial state before the quench, the atoms are evaporatively cooled to degeneracy in the two adjacent box traps with the DW barrier between the wells still low enough to allow tunneling. With this, we realize a pair of quasi-1D superfluids with high phase coherence between the clouds. The cooling procedure is the same as in~\cite{Schweigler2017} to assure that we prepare a thermal equilibrium state. When the evaporation ends the gases have a typical linear density of $\sim\SI{70}{\per\micro\meter}$, which corresponds to 2300 to 4800 atoms in each well depending on the box size. The tunnel coupling $J$ between the wells is chosen such that the phase of the two superfluids locks with a coherence factor $\cohfact$ from 0.8 to 0.95 ($J/2\pi = \;$\SIrange[range-units = single]{1}{3.5}{\hertz}).

\begin{figure}[t]
  \centering
  \includegraphics[width=0.55\textwidth]{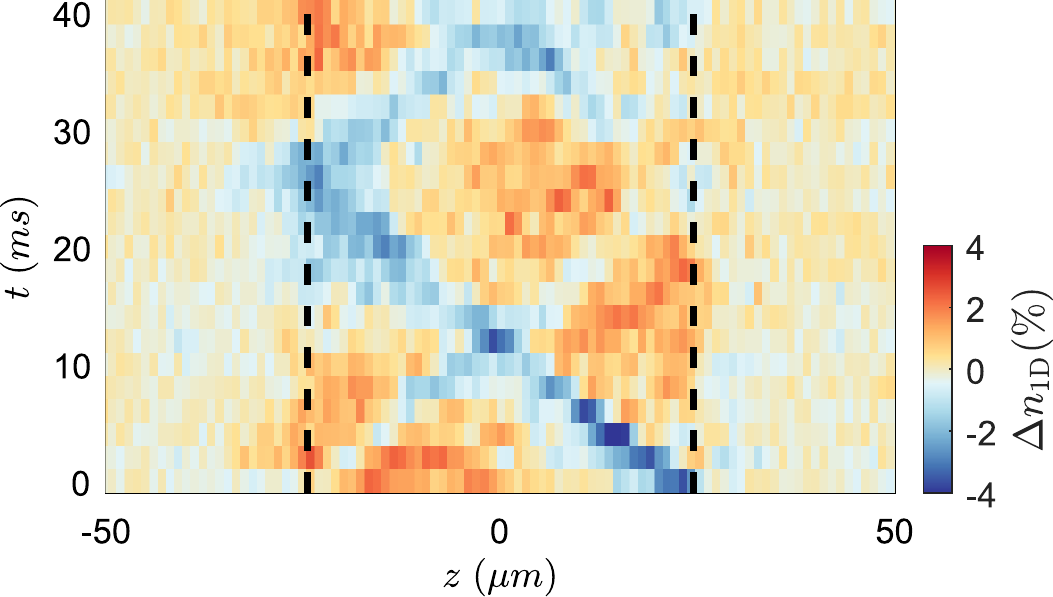}
  \caption{Density perturbations caused by the decoupling process. The density profile along the box confinement is measured by absorption imaging after \SI{2}{\milli\second} of free expansion. Shown is the deviation from the temporal profile mean $\overline{n_\mathrm{1D}}$ relative to the maximum density in percent 
$\Delta n_\mathrm{1D}(z) = (n_\mathrm{1D}(z) - \overline{n_\mathrm{1D}}(z))/\mathrm{max}(\overline{n_\mathrm{1D}}(z))$.
At the point of decoupling at $t=0$ density waves are launched from the box walls (dashed lines) and travel ballistically through the gas.
}
  \label{fig:density_waves}
\end{figure}

The two wells are then decoupled by increasing the amplitude of the RF magnetic field by \SI{13}{\percent} in $\sim$\SI{1.9}{\milli\second}. All evolution times given are with respect to the middle of this ramp. During the decoupling process the magnetic trap is slightly deformed. Specifically, the trap frequency in the tightly confined direction of each well increases by \SI{4}{\percent}, leading to a small interaction quench at the time of decoupling. As this process changes the equilibrium density profile abruptly, density waves of low amplitude are emitted from the walls, as seen in \fig{density_waves}. 
Additionally, a minor shift in the magnetic longitudinal potential causes the observed asymmetry. 
These waves are common to both superfluids and therefore do not affect the fluctuations in the relative degrees of freedom exhibiting the recurrences. Simulations of the Gross-Pitaevskii equation (GPE) (see \ref{SGPE_sect}) confirm that such waves can result from the change in trap frequency and qualitatively agree with the observed pattern. No influence of such perturbation on the recurrent behavior reported in the main text was observed in the simulations.

\subsection{Measurement of the relative phase}
The spatially resolved relative phase $\varphi(z)$ between the two superfluids is extracted form absorption images of the matter-wave interference pattern after \SI{15.6}{\milli\second} of time-of-flight (TOF) expansion~\cite{AduSmith2013a}. Schematically explained in \fig{fig_1} of the main text, each pixel column of the recorded absorption image is separately fitted by a sinusoidally modulated Gaussian~\cite{Langen2013}. The phase of that fitted sinusoidal modulation gives $\varphi(z)$ modulo $2\pi$. As we are only interested in the cosine of phase differences this ambiguity does not influence our analysis. 

For each point in time at which the evolution after the quench is measured the experiment is repeated 200 to 450 times. The phase correlation function $C(\bar z)$ is calculated by averaging $\mathrm{cos}(\varphi(z) - \varphi(z'))$ over all realizations and all points within the central region of the clouds that fulfill $\bar z = z - z'$ .



%% file: som_sec2_data_analysis.tex
\section{Data analysis}
\label{data_analysis_sect}

\subsection{Temperature measurement}

\begin{figure}[t]
  \centering
  \includegraphics[width=0.7\textwidth]{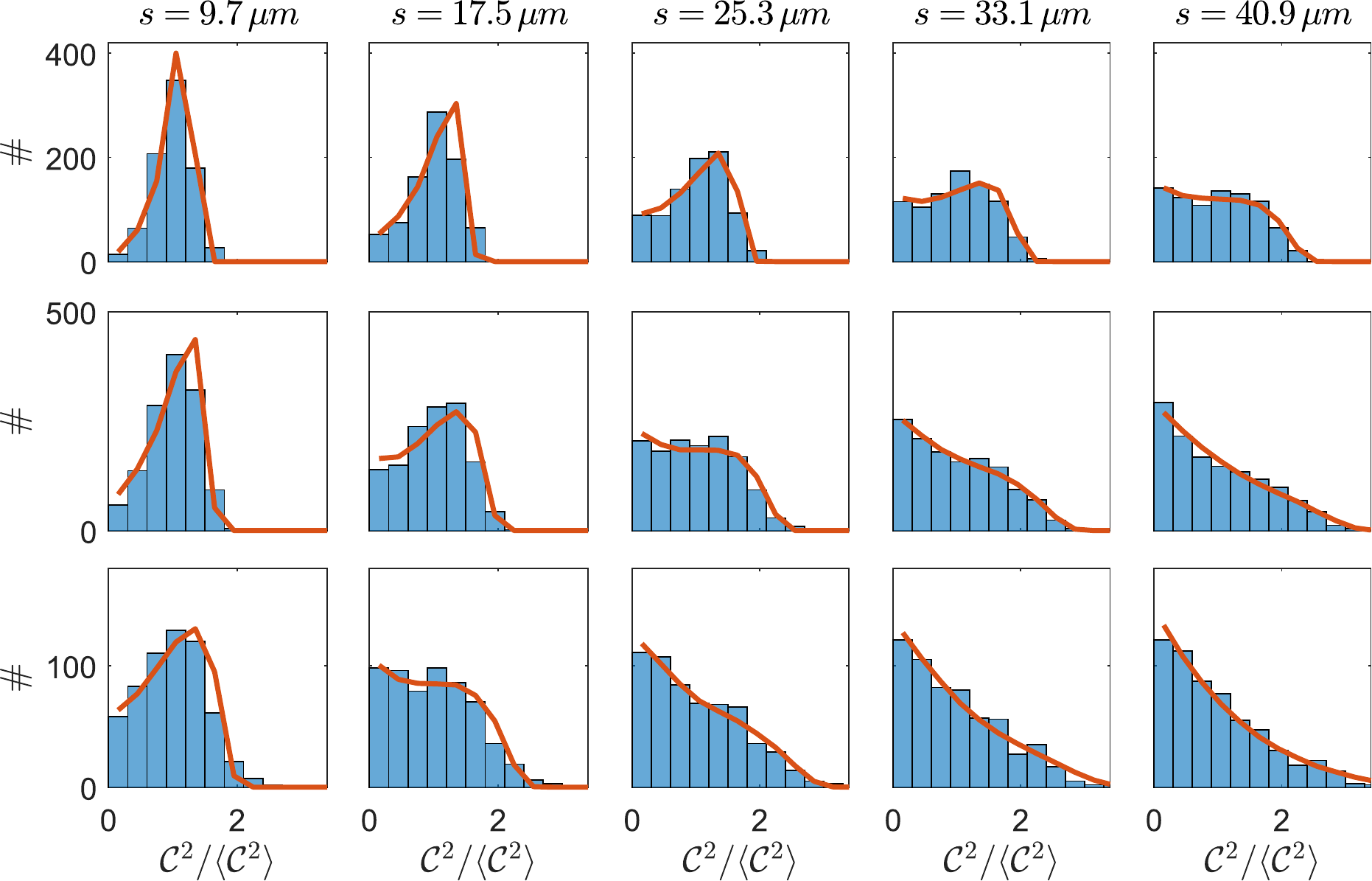}
  \caption{Distribution of the normalized squared measured contrast $\Cnorm$ (histograms) for three data sets and the respective fits (red solid lines) giving the effective temperatures $\Teff =\;$\SIlist[list-units = single,list-final-separator = {, \text{and} }]{43;72;100}{\nano\kelvin} (top to bottom). For each data set, the distributions are extracted from interference patterns integrated over five different lengths $s$.
}
  \label{fig:fdf_fits}
\end{figure}

The temperature of the initial state is measured by comparing the magnitude of the longitudinal density speckle patterns forming in the time-of-flight (TOF) expansion to simulations~\cite{Manz2010,Imambekov2009}. After decoupling, the effective temperature $\Teff$ of the relative degrees of freedom is extracted from the completely dephased state around $t=\trec/2$. It is inferred from the full distribution function of measured contrasts of the integrated interference patterns~\cite{Polkovnikov2006,gritsev2006,Hofferberth2008}. Specifically, the normalized squared contrast $\Cnorm$ distributions for different integration lengths $s$ are compared to results form simulated phase profiles~\cite{Stimming2010} in a $\chi^2$ fit. \Fig{fdf_fits} shows the results of such fits for three of the data points in \fig{fig_4} of the main text.

Even though the measurements are well described by thermal distributions the temperatures extracted from the relative degrees of freedom are treated as effective temperatures as the system is not fully in thermal equilibrium (see \ref{init_state_splitting}). Furthermore, these effective temperatures are consistently higher (up to \SI{30}{\nano\kelvin}) than the temperatures extracted for the initial state, indicating that the decoupling process adds additional energy to the system. For the analysis of the effect of temperature on the recurrence damping and the comparison to simulations we therefore use the effective temperature $\Teff$ of the relative degrees of freedom.

\subsection{Recurrence fit}
\label{rec_fit}

The recurrence time and height presented in \fig{fig_3}B and \ref{fig:fig_4} of the main text are extracted from fitting the time evolution of $C(t,\zc=\SI{27.3}{\micro\meter})$ with a Gaussian function. The distance $\zc$ is chosen such that it is long enough that the equilibrium phase correlations are low to facilitate a high recurrence visibility while being still much smaller than the system size even for the shortest box lengths. The applicability of a Gaussian fit is discussed in~\ref{rec_phase_coh}.

\begin{figure}[t]
  \centering
  \includegraphics[width=0.85\textwidth]{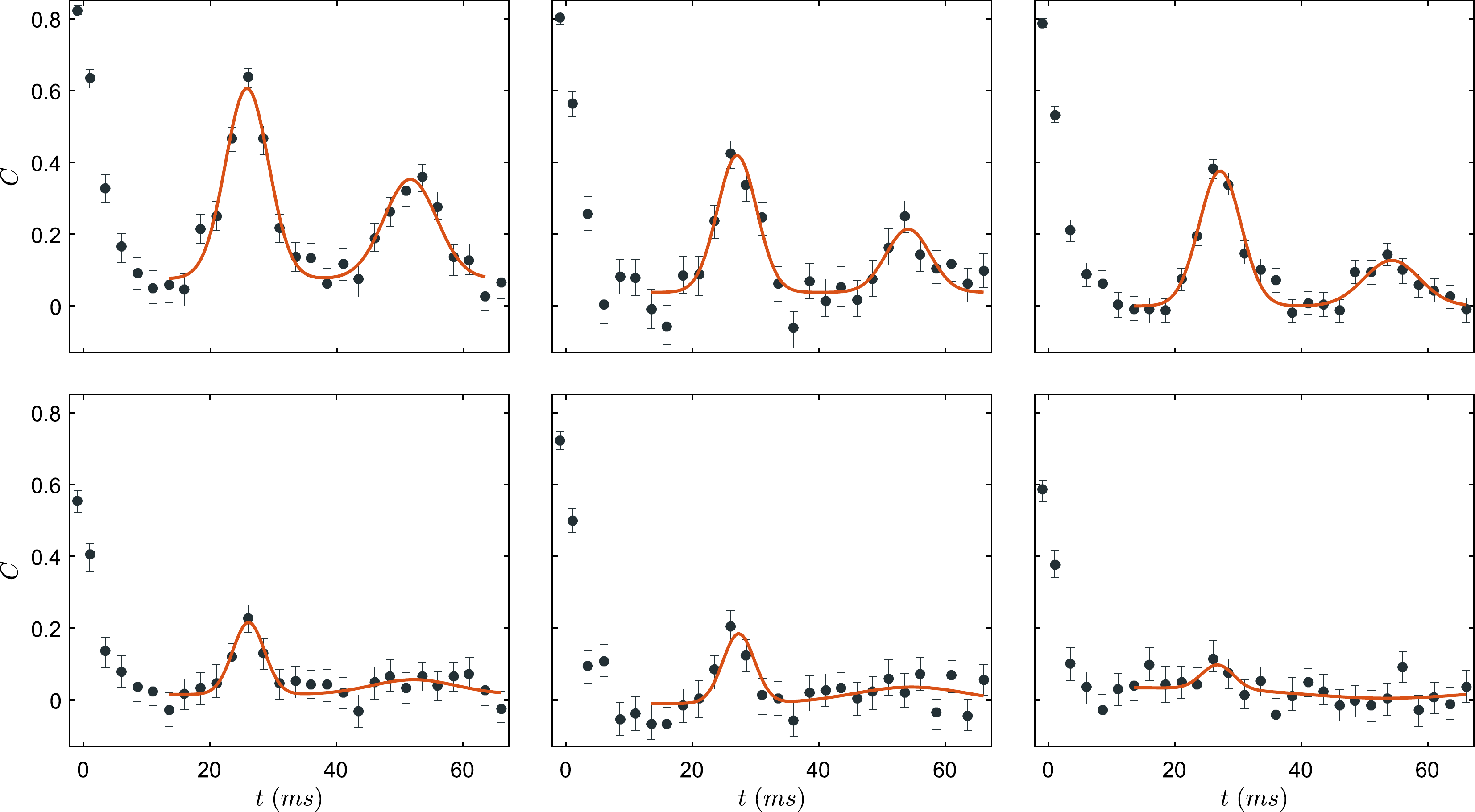}
  \caption{Temporal evolution of the phase correlation function at $\zc = \SI{27.3}{\micro\meter}$ (black circles) for all data sets presented in \fig{fig_4} of the main text. 
The corresponding effective temperatures are $\Teff =\;$\SIlist[list-units = single,list-final-separator = {, \text{and} }]{43;70;72}{\nano\kelvin}	(top row, left to right) and $\Teff =\;$\SIlist[list-units = single,list-final-separator = {, \text{and} }]{79;85;100}{\nano\kelvin}	(bottom row, left to right).
Fits to the first and second recurrence are shown in red. For the data set with the highest temperature the recurrences are nearly completely damped. However, the fit still provides a measure of excess correlations above the base line $\Cbase$ at the recurrence time 
}
  \label{fig:rec_fit}
\end{figure}

To extract the recurrence times the first and second recurrence are fitted independently. The results of these fits for measurements with different box lengths are presented in \fig{fig_3}B of the main text.

For the height fit of the hot data sets presented in in \fig{fig_4} of the main text we perform a combined fit of both recurrences with the restricted double Gaussian

\begin{equation}
	f(t) = \Cbase + \sum_{i=1}^{2} C_\mathrm{fit}^{(i)} \, e^{-\frac{(t - i\cdot\trec)^2}{2\sigma_i^2}}, \nonumber
\end{equation}

\noindent where $i=1,2$ is the recurrence index. Only the amplitudes $C_\mathrm{fit}^{(i)}$ and widths $\sigma_i$ of both peaks are varied in this fit while the base line $\Cbase$ and the peak positions are fixed. The positions are set by $\trec$ obtained from the initial unrestricted fit (only for the hottest data point in \fig{fig_4} of the main text the value of $\trec$ is provided by a Luttinger liquid calculation as the unrestricted fit is not stable due to the nearly vanishing recurrence signal). $\Cbase$ is fixed to the averaged values of $C(t,z_c)$ in between the recurrences at $t = \frac{\trec}{2},\frac{3\trec}{2}$ and $\frac{5\trec}{2}$. Examples of such a fits are presented in \fig{rec_fit}.

From this fit we define the recurrence height as   

\begin{equation}
	\Hrec^{(i)} = \frac{C_\mathrm{fit}^{(i)} - \Cbase}{C(0) - \Cbase}, \nonumber
\end{equation}

\noindent with $C(0)$ being the initial correlations before decoupling. This measure is one for a perfect recurrence and zero if no excess correlation beyond the base level $\Cbase$ occurs.

The confidence intervals given in the main text for the recurrence time and height were obtained using a bootstrap (bias corrected and accelerated method), specifically its implementation 'bootci' in Matlab.



%% file: som_sec3_theory.tex
\section{Theoretical description}

\subsection{Low-energy effective theory}
We consider an ultra cold gas of $^{87}$Rb atoms in a DW potential on an atom chip.
The Hamiltonian of the system is given by
\begin{equation}
 H = \int \mathrm{d}{\bf{r}}~ \Psi^{\dagger}({\bf{r}}) ~\left[ -\frac{\hbar^2}{2m} \nabla^2 + V({\bf{r}}) - \mu + g_{3\mathrm{D}} \Psi^{\dagger}({\bf{r}}) \Psi({\bf{r}}) \right]~\Psi({\bf{r}}) ~\mathrm{,}
\end{equation}
where $m$ is the atomic mass, $g_{3\mathrm{D}}=4\pi \hbar^2 a_s / m$ is the 3D interaction constant in the limit of s-wave scattering, $a_s$ is the s-wave scattering length, $\hbar$ is the reduced Planck constant, and 
$\mu$ is the chemical potential.
The field operators fulfill bosonic commutation relations $[\Psi({\bf{r}}), \Psi^{\dagger}({\bf{r'}})] = \delta({\bf{r}}-{\bf{r'}})$. 
The external trapping potential $V({\bf{r}})$ is decomposed into $V({\bf{r}}) = V_{\mathrm{DW}}(x) V_{h}(y) U(z)$, where $V_{\mathrm{DW}}(x)$ is the DW potential, $V_{\mathrm{h}}(y) = m \omega_{\perp}^2 y^2 / 2$ is 
a tight harmonic confinement ($\omega_{\perp} \simeq \angfreq{1.4}{\kilo}$), and $U(z)$ is the box potential in the longitudinal direction.
Around the two minima, $V_{\mathrm{DW}}$ is well approximated by a harmonic potential with the trap frequency $\omega_{\perp}$.
Both condensates, located in the left and right minimum of the DW potential, fulfill the 1D condition, $\mu, k_{\mathrm{b}} T < \hbar \omega_{\perp}$, and therefore dynamics along the radial $(x,y)$ directions are frozen out. 
However, tunneling trough the DW barrier couples the two superfluids. Usually each condensate is considered to be in the single particle ground state $\phi_0$ of the transverse harmonic confinement, 
which leads after integration to the effective 1D interaction constant
\begin{equation} \label{eq:bare_1d_coupling}
	g_\mathrm{1D} = g_{3\mathrm{D}} \int{|\phi_0(x,y)|^4\mathrm{d}x\mathrm{d}y} = 2\hbar a_s \omega_\perp, 
\end{equation}
with the integration over the kinetic and potential energy terms only leading to a constant energy shift of $\hbar\omega_\perp$.
Note that this holds only if $a_s$ is much smaller than the harmonic oscillator length $a_\perp = \sqrt{\hbar/m\omega_\perp}$ which is the case for our parameters~\cite{Olshanii1998}. 

However, the use of the single particle ground state $\phi_0$ relies on the assumption of low 1D densities $n_\mathrm{1D} a_s \ll 1$. A useful approach to go beyond this assumption and to take transverse interactions into account 
is developed in~\cite{Salasnich2002}. The transverse wave function is assumed to be a locally broadened Gaussian with $\sigma(z) = a_\perp \sqrt{1 + 2 a_s n_\mathrm{1D}(z)}$ calculated by a variational approach.
Inserting this broadened transverse wave function into the 3D Hamiltonian and integrating over the two tightly confined directions $x,y$ leads to
\begin{align} \label{eq:Hcoupledcondensates}
H = \sum_{j=1}^2 &\int \mathrm{d}z ~ \psi^{\dagger}_j(z) \left[ -\frac{\hbar^2}{2m} \left( \frac{\partial}{\partial z } \right)^2
                                                              + {U}(z) - \mu + \hbar \omega_{\perp} \sqrt{1+2 a_s |\psi_j(z)|^2} \right] \psi_j(z) \notag \\
       &- \hbar J  \int \mathrm{d}z ~ \left[ \psi^{\dagger}_1(z) \psi_2(z) + \psi^{\dagger}_2(z) \psi_1(z) \right] ~,
\end{align}
with $j=1,2$ being the index of the wells.
Here we neglected the influence of the tunnel coupling to the broadening of the radial wave function. This is justified for two reasons: Firstly, in thermal equilibrium the tunneling energy is much smaller than the interaction 
energy in a single condensate and therefore only leads to minor corrections of the transverse width. Secondly, $J$ is an effective parameter in our model determined from experimental measurements and not calculated ab initio from 
the shape of the DW potential. Therefore, a renormalization of $J$ is irrelevant as we are only interested in the stationary thermal state of the coupled system. 
Note that for the evolution following the quench to the decoupled system, the above approximations are irrelevant as $J=0$.

In order to derive a low-energy effective theory we can follow the same procedure as ~\cite{Mora2003} and express the 1D wave function in terms of density and phase fluctuations $\psi_j = e^{i\theta_j}\sqrt{n_\mathrm{1D} + \delta\rho_j}$,
with canonical commutators $[\delta {\rho}_j(z), \theta_{j'}(z')] = \mathrm{i} \, \delta_{j j'}\delta(z-z')$. Expanding in powers of the small density fluctuations $\delta\rho_j$ and phase gradients $\partial_z \theta_j$ to quadratic order 
separates the Hamiltonian (\ref{eq:Hcoupledcondensates}) into a weakly coupled sum $H = H_{\mathrm{c}} + H_{\mathrm{r}} + V_{\mathrm{c},\mathrm{r}}$ of common (c) and relative (r) degrees of freedom, defined as
\begin{align}
&\delta {\rho}_c(z) = \delta {\rho}_1(z) + \delta {\rho}_2(z)  \hspace{0.01\textwidth}~\mathrm{,} &&{\varphi}_c(z) = \frac{1}{2} [{\theta}_1(z) + {\theta}_2(z)]~\mathrm{,} \\
&\delta {\rho}_r(z) = \frac{1}{2} [\delta {\rho}_1(z) - \delta {\rho}_2(z)]~\mathrm{,} \hspace{0.01\textwidth} &&{\varphi}_r(z) = {\theta}_1(z) - {\theta}_2(z) ~\mathrm{.}
\end{align}
It was experimentally shown in \cite{Schweigler2017}, that the relative degrees of freedom relevant for our analysis are in thermal equilibrium well described solely by $H_{\mathrm{r}}$ for a wide range of parameters. For a general coupling strength $J$, the Hamiltonian $H_{\mathrm{r}}$
can be approximated by the sine-Grodon model. However, for the quench performed in the experiment we are only interested in the two limiting cases of very strong and vanishing tunnel coupling. In the former the large tunnel coupling leads to strong phase-locking
between the two condensates, and therefore an overall small phase ${\varphi}_r$. This allows to further approximate $H_{\mathrm{r}}$, taking into account only terms which are quadratic in the phase field itself. We then find the initial strongly coupled system to be described by the Hamiltonian
 \begin{align}
H = \int dz \biggl[ 
\frac{\hbar^2}{4m n_\mathrm{1D}} \biggl( \frac{\partial {\delta \rho}}{\partial z}\biggr)^2  + g\delta {\rho}^2 + 
\frac{\hbar^2 n_\mathrm{1D}}{4m} \biggl( \frac{\partial {\varphi}}{\partial z}\biggr)^2 -\, \hbar J n_\mathrm{1D} \varphi^2 \biggr] \label{eq:fullSecondOrder} \, ,
\end{align}
where we omitted the subscript `$r$', as we do in the following and the main text. Further, we defined the effective interaction constant 
\begin{align}
g = \hbar \omega_\perp a_s \frac{2 + 3 a_s n_\mathrm{1D}}{(1 + 2 a_s n_\mathrm{1D})^{3/2}} - \frac{\hbar J}{n_{\mathrm{1D}}} \label{eq:1dinteractionNPSE} \, .
\end{align}
In the case of vanishing tunnel coupling the system is described by the same equations with $J$ set to zero. 
Further neglecting the first term of the Hamiltonian given by the derivative of density fluctuations, and therefore at the energies considered highly suppressed, restricts the spectrum of (\ref{eq:fullSecondOrder}) to the linear regime. 
In case of completely decoupled superfluids this leads to the well known Luttinger liquid Hamiltonian. The quadratic Hamiltonian (\ref{eq:fullSecondOrder}) can be exactly diagonalized using a Bogoliubov transformation. 
Representing the density and phase fluctuations in terms of elementary excitations
\begin{align} 
 \delta \rho &= \frac{1}{\sqrt{2 n_{1 \mathrm{D}}}} \sum_m \bigg[ \mathrm{i} f_{m}^{-}(z)~\mathrm{e}^{-\mathrm{i} \omega_m t}~b_m + \mathrm{H.c.} \bigg] \label{eq:expansionBdG_dens} \\
 \varphi &= \frac{1}{\sqrt{2 n_{1 \mathrm{D}}}} \sum_m \bigg[ f_{m}^{+}(z)~\mathrm{e}^{-\mathrm{i} \omega_m t}~b_m + \mathrm{H.c.} \bigg] \label{eq:expansionBdG_phase} ~ ,
\end{align}
leads to the following equations for the eigenvalues $\epsilon_m = \hbar \omega_m$ and eigenfunctions $f^{\pm}_m$ of the excitations:
\begin{align}
\epsilon_m f_{m}^{+}(z)  &= \bigg[ -\frac{\hbar^2}{2m} \left(\frac{\partial}{\partial z} \right)^2 + U(z) + 3 g n_{1 \mathrm{D}} - \mu_0 + 2 \hbar J \bigg] f_{m}^{-}(z) \\
\epsilon_m f_{m}^{-}(z)  &= \bigg[ -\frac{\hbar^2}{2m} \left(\frac{\partial}{\partial z} \right)^2 + U(z) + ~g n_{1 \mathrm{D}} - \mu_0 + 2 \hbar J \bigg] f_{m}^{+}(z) ~\text{.} 
\end{align}
Here $\mu_0$ is the unshifted chemical potential for vanishing tunnel coupling. The commutation relation $[\delta {\rho}(z), \varphi(z')] = \mathrm{i} \, \delta(z-z')$ ensures the normalization of the mode functions
\begin{align}
 \frac{1}{2} \int \mathrm{d}z ~ \bigg[ (f_{m}^{+})^{*} (f_{m}^{-}) + (f_{m}^{-})^{*} (f_{m}^{+}) \bigg] = 1 ~\text{.}
\end{align}
We consider an idealized box shaped potential $U(z) = U_0 [\Theta(-z) + \Theta(z-L)]$, neglecting the influence of the a finite wall steepness and the deformation of the background density $n_{1 \mathrm{D}}(z)$ close to the boundary 
(on a length scale of the order of the healing length $\xi_h$ of a single condensate). Considering the limit $U_0 \to \infty$ the particle flux at the boundary has to vanish, being equivalent to Neumann boundary conditions $\partial_z \varphi |_{z=0,L} = 0$.
In this case, we can give the solutions of the above eigenvalue problem
\begin{align}
  f^{\pm}_{k,J} &= \left( \frac{\epsilon_{k,J}}{E_k+2 \hbar J} \right)^{\pm 1/2} \frac{1}{\sqrt{L}} \operatorname{cos}(k z) \\
  \epsilon_{k,J} &= \sqrt{(E_k + 2 \hbar J) \, ( E_k + 2 \hbar J + 2 \mu_0 )} ~ , \label{eq:dispersionBdG}
\end{align}
where $E_k = \hbar^2 k^2 / 2m$ and the discrete states are labeled by their momentum $k= \pi n / L$ with $n$ being an integer number.

\subsection{Phase correlation functions}
The observable investigated in the experiment is the full two-point correlation function, given by
\begin{equation}\label{eq:PCF_fields}
 C(z_1,z_2,t) = \frac{\langle \psi_1(z_1,t) \psi_{2}^{\dagger}(z_1,t) \psi_{1}^{\dagger}(z_2,t) \psi_{2}(z_2,t) \rangle}{\langle |\psi_1(z_1,t)|^2 \rangle \langle |\psi_2(z_2,t)|^2 \rangle} ~\mathrm{.}
\end{equation}
In the limit of small density fluctuations this can be approximated as
\begin{align} \label{eq:PCF}
 C(z_1,z_2,t) \approx \langle \operatorname{cos}\left[ \varphi(z_1,t)-\varphi(z_2,t) \right] \rangle = \mathrm{exp} \left( -\frac{1}{2} \langle [\varphi(z_1,t)-\varphi(z_2,t)]^2 \rangle \right) ~,
\end{align}
where for the second equality we assumed Gaussian phase fluctuations.
Calculating the explicit expression for the variance of phase differences $\Delta\varphi(z_1,z_2,t) = \varphi(z_1,t)-\varphi(z_2,t)$ from (\ref{eq:expansionBdG_phase}), we obtain
\begin{align}\label{eq:PCF_full}
 \langle \Delta\varphi(z_1,z_2,t)^2 \rangle = \frac{1}{2 n_{1\mathrm{D}}} \sum_{\epsilon_{k,J}>0} \Big\{ &~|f_{k,J}^{+}(z_1) - f_{k,J}^{+}(z_2)|^2 ~\Big(2 \langle b_{k}^{\dagger} b_k \rangle +1 \Big) \notag \\
	&+ \Big( \mathrm{e}^{-2 \mathrm{i} \omega_{k,J} t} ~[f_{k,J}^{+}(z_1) - f_{k,J}^{+}(z_2)]^2 ~\langle b_k b_k \rangle + \mathrm{H.c.} \Big) \Big\} ~ ,
\end{align}
with the mode frequency $\omega_{k,J} = \epsilon_{k,J} / \hbar$ and neglecting correlations between different $k$-modes which will be justified below.

\subsection{Initial state and instantaneous splitting}
\label{init_state_splitting}

The system is prepared in a thermal state of the strongly tunnel coupled DW. Evaluating the expectation values of the quasi-particle operators for a thermal state, Eq.~(\ref{eq:PCF_full}) reduces in case of a box shaped potential to 
\begin{align}
 \langle \Delta\varphi(z_1,z_2,t)^2 \rangle = \frac{1}{L n_{1\mathrm{D}}} \sum_{\epsilon_{k,J}>0} ~\frac{\epsilon_{k,J}}{E_k+2 \hbar J} \left[ \operatorname{cos}(k z_1) - \operatorname{cos}(k z_2) \right]^2 \left( \frac{k_{\mathrm{b}} T}{\epsilon_{k,J}} + \frac{1}{2} \right) ~ ,
\end{align}
where the only non-vanishing contributions to the sum are due to the quasi-particle occupation $n_k = \langle b_{k}^{\dagger} b_k \rangle$. For the temperatures considered we can approximate the Bose-Einstein distribution by  $n_k = k_{\mathrm{b}} T / \epsilon_{k,J}$ 
and will neglect in the following the minor $T=0$ quantum contributions.

At the time $t=0$ the coupling $J$ is quenched to zero, completely decoupling the two condensates. The system posterior to the quench evolves according to the uncoupled Hamiltonian, with the initial conditions defined by the projection of the 
coupled thermal state onto the new basis. By inversion of Eq.~(\ref{eq:expansionBdG_dens},\ref{eq:expansionBdG_phase}) we get
\begin{align}
 n_k = \langle b_{k}^{\dagger} b_k \rangle &= \frac{k_{\mathrm{b}} T}{2 \epsilon_{k,0}} \left( \frac{E_k}{E_k + 2 \hbar J} + \frac{E_k + 2 \hbar J}{E_k + 2 \hbar J + 2 \mu_0} \right) \\
 m_k = \langle b_{k}^{(\dagger)} b_{k}^{(\dagger)} \rangle &= \frac{k_{\mathrm{b}} T}{2 \epsilon_{k,0}} \left( \frac{E_k}{E_k + 2 \hbar J} - \frac{E_k + 2 \hbar J}{E_k + 2 \hbar J + 2 \mu_0} \right) ~ .
\end{align}
Here $\epsilon_{k,0}$ is the dispersion relation for the uncoupled system, see Eq.~(\ref{eq:dispersionBdG}) with $J=0$. 
Due to the orthogonality of the mode functions, all other expectation values vanish identically. Therefore the time-dependent phase correlation function Eq.~(\ref{eq:PCF_full}) is determined by
\begin{align} \label{eq:PCFpostquench}
 \langle \Delta\varphi(z_1,z_2,t)^2 \rangle = \frac{1}{n_{1\mathrm{D}}} \sum_{\epsilon_{k,0}>0} ~|f_{k,0}^{+}(z_1) - f_{k,0}^{+}(z_2)|^2 ~ \left[ n_k+m_k - 2 m_k \operatorname{sin}^2(\omega_{k,0} t) \right] ~ .
\end{align}
In the short time evolution decoherence proceeds through light-cone-like dephasing of quasi-particle modes. In the thermodynamic limit ($L \to \infty$, $n_{1\mathrm{D}}=\mathrm{const.}$) the system would relax to the steady state of this 
low-energy effective theory when looking at local observables. In the long time limit, higher-order corrections to the Hamiltonian would lead to the complete relaxation of the system. No recurrence of phase coherence would be present, as the dispersion relation
becomes denser with increasing system size, until in the thermodynamic limit it becomes a continuous function of $k$. 

\subsection{Recurrence of phase coherence}
\label{rec_phase_coh}

\begin{figure}[t]
  \centering
  \includegraphics[width=0.45\textwidth]{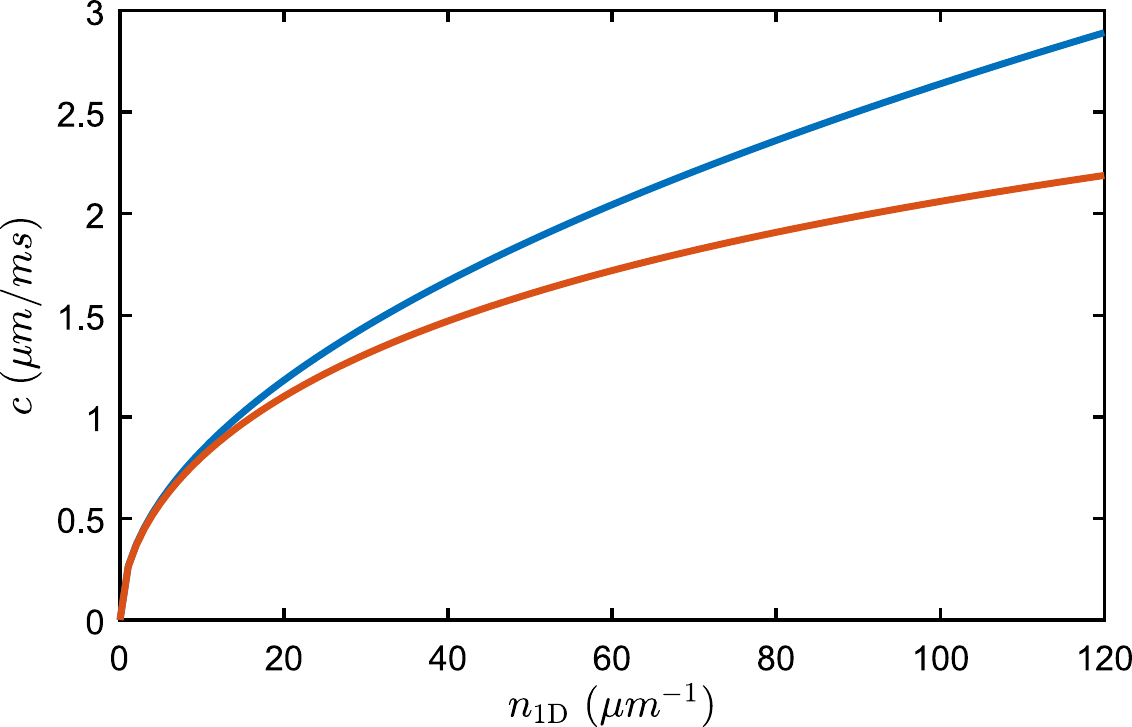}
  \caption{Speed of sound of a 1D superfluid trapped in a transverse harmonic confinement with $\omega_\perp = 2\pi \cdot 1.45\,$kHz. The bare speed of sound $c_0$ is plotted in blue. Transverse interactions slow down the dynamics and lead to the effective speed of sound $c$ plotted in red. 
}
  \label{fig:speed_of_sound}
\end{figure}

In our finite size system however, the lowest lying modes are discrete, leading to a finite lowest eigenfrequency of the system, and therefore 
a finite oscillation period. As contributions of high-energy modes to the phase correlation function are greatly suppressed, we can linearize the dispersion relation (\ref{eq:dispersionBdG}) for vanishing tunnel coupling, leading to 
$\epsilon_{k,0} \approx \hbar c k$. The time evolution of the phase correlation function is therefore dominated by a set of commensurate mode frequencies, which results in the earliest possible time for a full recurrence 
given by $2\pi / \operatorname{min}(\Delta\omega) = 2 L / c$. The earlier recurrence of phase coherence for the inverted state at the time $\trec = L / c$ reported in the main text is apperent from Eq.~(\ref{eq:PCFpostquench}).  
Explicitly calculating the speed of sound
\begin{equation}
	c = \sqrt{\frac{g n_\mathrm{1D}}{m}} = c_0 \sqrt{\frac{1}{2} \frac{2 + 3 a_s n_{1\mathrm{D}}}{(1 + 2 a_s n_\mathrm{1D})^{3/2}}} ~,
\end{equation}
one finds a non-trivial density dependence due to the broadening of the transverse wave function, and therefore a shift as compared to the bare value $c_0 = \sqrt{\frac{g_{1\mathrm{D}} n_\mathrm{1D}}{m}}$ 
determined by Eq.~(\ref{eq:bare_1d_coupling}). \Fig{speed_of_sound} shows that transverse interactions slow down the dynamics. We see that for a typical central density of the measurements discussed in this manuscript 
($n_\mathrm{1D} = \SI{70}{\per\micro\meter}$), the deviation from the bare speed of sound is about 20\%, substantially changing the expected recurrence times.   

Linearizing Eq.~(\ref{eq:PCFpostquench}) in time near $\trec$ shows that the time evolution of the phase correlation function Eq.~(\ref{eq:PCF}) for fixed spatial coordinates $z_1$,$z_2$ is given by the product of Gaussian functions with 
$k$-dependent widths. As a product of Gaussian functions is again a Gaussian, the temporal shape of the phase correlation functions near a recurrence is Gaussian as well. Its variance is given by the sum of the original variances meaning that the width of the recurrence signal in the correlation function decreases with temperature.

%% file: som_sec4_numerics.tex
\section{Numerical simulations}
\label{num_sim}

\subsection{Solution of the Luttinger liquid model}

In order to distinguish the damping originating from imperfections of the experimental box trap or fluctuations in the initial conditions from genuine higher order corrections to the low-energy effective theory we numerically solve the quadratic Hamiltonian (\ref{eq:fullSecondOrder}) 
within the linear regime $\epsilon_{k,0} = \hbar c k$. This is equivalent to neglecting the first term in the Hamiltonian (\ref{eq:fullSecondOrder}), resulting in the Luttinger liquid model. The numerical solution is obtained as follows: 
We consider the spatially discretized system, where $\vec{\varphi}$ and $\vec{\delta\rho}$ are vectors of the discrete lattice model used to approximate the continuous quantum field theory. 
Since density and phase fluctuations are completely decoupled the density matrix for a thermal state can be written in matrix form as
\begin{align}
 \rho = \frac{1}{Z} \operatorname{exp}\left[-\beta H \right] = \frac{1}{Z} \operatorname{exp}\left[-\beta \begin{pmatrix} \vec{\varphi} ~ \vec{\delta\rho} \end{pmatrix}^{\mathrm{T}} \begin{pmatrix} K & 0 \\ 0 & L \end{pmatrix} \begin{pmatrix} \vec{\varphi} \\ \vec{\delta\rho} \end{pmatrix} \right] ~,
\end{align}
where $\beta=1 / k_{\mathrm{B}} T$ is the inverse temperature, $Z = \operatorname{Tr} \left[ \operatorname{exp}(-\beta H) \right]$ is the partition function, and $K, L$ are matrices defined by the discretization of the Hamiltonian $H$. 
The density matrix is given by a Gaussian multivariate distribution and density and phase variances can be numerically determined in the classical field approximation by inverting the matrices $K$ and $L$. 
The subsequent time evolution can be readily calculated for this exactly solvable model. We find two contributions to the damping of the recurrence height in this model. 
Firstly, the incorporation of the experimental trapping potential 
determining the background density profile $n_{1\mathrm{D}}(z)$ slightly deforms the dispersion relation, causing the mode frequencies not to be perfectly commensurate. Secondly, taking into account fluctuations of the total 
atom number and the imbalance between the wells, the model exhibits damping caused by the averaging over realizations with a different speeds of sound and therefore different recurrence times $\trec$. 
However, as shown in the main text, these effects do not explain the large damping observed in the experiment. Therefore, higher-order corrections beyond the harmonic approximation, faithfully incorporated in the GPE formalism, 
need to be considered.

\subsection{Stochastic Gross-Piteavskii equation}
\label{SGPE_sect}

The initial thermal state for the numerical simulations of the coupled 1D system (\ref{eq:Hcoupledcondensates}) is created using the stochastic (non-polynomial) Gross-Pitaevskii equation (SGPE) \cite{Stoof1999,Davis2001,Gardiner2002}. 
The SGPE describes the dynamics of the coherent region $\psi_j$, for the individual condensates $j=1,2$, while being in contact with a thermal cloud via the Langevin equation
\begin{equation}\label{eq:SGPE}
 \mathrm{i}\hbar \partial_t \begin{pmatrix} \psi_1 \\ \psi_2 \end{pmatrix} = \Big[ H_{\mathrm{cGP}} - \mathrm{i}\mathcal{R}(z,t) \Big] \begin{pmatrix} \psi_1 \\ \psi_2 \end{pmatrix} + \begin{pmatrix} \eta_1 \\ \eta_2 \end{pmatrix} ~\mathrm{.}
\end{equation}
Here $H_{\mathrm{cGP}}$ denotes the operator for the coupled 1D system given by
\begin{equation}
 H_{\mathrm{cGP}} = \begin{pmatrix} H_{\mathrm{GP}}(\psi_1) & -\hbar J \\ -\hbar J & H_{\mathrm{GP}}(\psi_2) \end{pmatrix} ~\mathrm{,}
\end{equation}
where we defined the Gross-Pitaevskii operator for a single condensate as
\begin{equation} \label{eq:HamiltonianNPSE}
 H_{\mathrm{GP}}(\psi_j) = - \frac{\hbar^2}{2m} \partial_z^2 + U(z) - \mu + \hbar \omega_{\perp} \sqrt{1+2 a_s |\psi_j(z)|^2} ~\mathrm{.}
\end{equation}
For the comparison of the recurrence height with the experiment, we consider fluctuations of the total atom number determined from the experimental data.
We choose the chemical potential $\mu$ to be the same leading to a small spread of imbalance between the wells caused by thermal fluctuations.
Apart from the unitary evolution of the coupled semi-classical fields, Eq.~(\ref{eq:SGPE}) features two additional 
effects: The dissipative term $\mathcal{R}$ represents the particle transfer between the coherent region (low-energy modes) and the thermal reservoir (high-energy modes). 
The complex fields $\eta_j \equiv \eta_j(z,t)$ are independent Gaussian white noise, representing the random nature of incoherent scattering within the system. They are completely determined by their second moments
$\langle \eta_i(z,t)\eta_j(z',t') \rangle = (\mathrm{i}/2) \hbar^2 \Sigma^{\mathrm{K}}(z,t)\delta(z-z')\delta(t-t')\delta_{ij}$. Their strength is given by the Keldysh self-energy $\Sigma^{\mathrm{K}}$,
which in general must be calculated consistently, taking into account the dynamics of the high-energy region (described by a quantum Boltzmann equation).

\begin{figure}[t]
  \centering
  \includegraphics[width=0.95\textwidth]{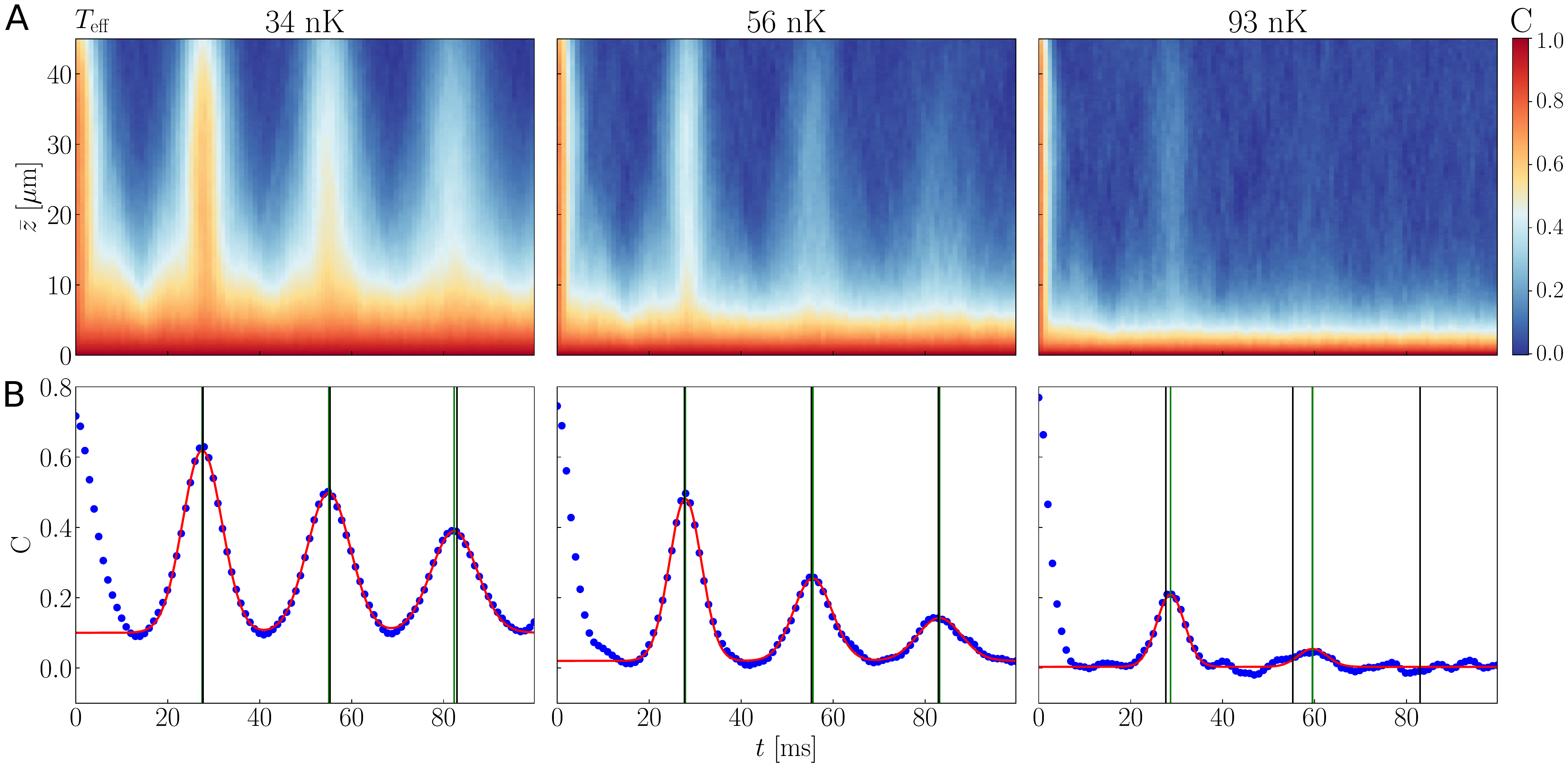}
  \caption{Results of the Gross-Pitaevskii simulations for three different temperatures \SIlist[list-units = single,list-final-separator = {, \text{and} }]{34;56;93}{\nano\kelvin} (left to right). (\textbf{A}) Time evolution of the phase correlation function $C(\bar{z},t)$ with the 
  amount of correlation depicted in color. The damping of revivals with temperature is clearly visible. (\textbf{B}) Temporal evolution for a cut at $\zc = \SI{27.3}{\micro\meter}$, taking into account the experimental resolution. 
  Results of the numerical simulation (blue dots) are well described by a Gaussian fit (red lines). The extracted revival time is depicted by the green vertical lines, and is in good accordance with the Luttinger liquid predictions for an ideal box taking into account the transverse broadening of the wave function. 
}
  \label{fig:GPE}
\end{figure}

In order to arrive at a numerically feasible theory, the high-energy region of the system is assumed to be in thermal equilibrium. Therefore, this region acts as a static heat bath.
As a consequence, the noise and the dissipative term in the SGPE are connected through the Bose-Einstein distribution, i.e.\ the fluctuation-dissipation relation. 
To further simplify the equations we consider the high-temperature (large-occupation) limit, and approximate the Bose-Einstein distribution with the Rayleigh-Jeans limit.
With the classical form of the fluctuation-dissipation relation, we can write the SGPE Eq.~\eqref{eq:SGPE} in the closed form
\begin{equation}\label{eq:SGPE2}
 \mathrm{i}\hbar \partial_t \begin{pmatrix} \psi_1 \\ \psi_2 \end{pmatrix} = [1 - \mathrm{i}\gamma(z,t)] H_{\mathrm{cGP}} \begin{pmatrix} \psi_1 \\ \psi_2 \end{pmatrix} + \begin{pmatrix} \eta_1 \\ \eta_2 \end{pmatrix} ~\mathrm{,}
\end{equation}
where $\gamma(z,t) \sim \beta \Sigma^{\mathrm{K}}(z,t)$, with the inverse temperature $\beta = 1/k_{\mathrm{B}} T$, and the fluctuations of the noise are given by 
$\langle \eta_i(z,t)\eta_j(z',t') \rangle = 2 \hbar \gamma(z,t) k_{\mathrm{B}} T \delta(z-z')\delta(t-t')\delta_{ij}$. Since we are only interested in creating a thermal state of the system, and not the time-dependent condensation 
dynamics, the dimensionless coupling $\gamma$ can be chosen arbitrarily as it merely rescales the condensation time. We choose a spatially constant rate $\gamma=0.05$. In order to avoid the excitation of 
topological defects during the condensation process the initial state is chosen with a finite occupation of the condensate and evolved until convergence to the thermal state is observed.

In the experiment the coupling $J$ is subsequently ramped down within a time of $\approx 1.9 \, \mathrm{ms}$, wherein the system completely decouples on a much shorter timescale. Therefore, in the numerical simulations, we consider an instantaneous 
quench of the coupling $J$ at the time $t=0$ from some finite value to zero. We further remove the contact with the heat bath and consider the unitary evolution of the system under the 
Hamiltonian (\ref{eq:HamiltonianNPSE}). 
The simulations are performed on a spatial grid of $N=1024$ points with a grid spacing of $a_{\mathrm{G}} = \SI{0.1}{\micro\meter}$ using a second-order time splitting Fourier pseudo-spectral method. 
Expectation values are calculated averaging over $2500$ independent realizations. The experimental imaging resolution is considered by convolving the relative phase $\varphi$ with a Gaussian of width $\sigma=\SI{3}{\micro\meter}$.
For this, the phase needs to be corrected to eliminate unphysical phase jumps $|\Delta \varphi_{i,i+1}|>\pi$ between neighboring points i and i+1 on the numerical grid. 
Without this correction, these unphysical jumps are smeared out, severely altering the phase correlations. The overall ambiguity of $2 \pi n$ with $n$ being an integer number is irrelevant for our analysis.

The analysis of the recurrence height is performed analogously to the experiment (see \ref{data_analysis_sect}). The effective temperature $\Teff$ in between the recurrences is extracted from the phase correlation function (\ref{eq:PCF})
by fitting the thermal predictions of a homogenous system to the central part of the cloud. As for the experimental data, the recurrence times and heights are extracted by fitting a Gaussian function to each recurrence for a slice through the phase correlation function at $\zc=\SI{27.3}{\micro\meter}$ 
(averaged over the experimental pixel size in image space of $\SI{2}{\micro\meter}$). In Fig.~(\ref{fig:GPE}) we show the time evolution of the phase correlation function $C(\bar{z},t)$ and the results of the Gaussian fits for different temperatures $\Teff$.

%% file: recurr_arxiv.bbl
\begin{thebibliography}{46}%
\makeatletter
\providecommand \@ifxundefined [1]{%
 \@ifx{#1\undefined}
}%
\providecommand \@ifnum [1]{%
 \ifnum #1\expandafter \@firstoftwo
 \else \expandafter \@secondoftwo
 \fi
}%
\providecommand \@ifx [1]{%
 \ifx #1\expandafter \@firstoftwo
 \else \expandafter \@secondoftwo
 \fi
}%
\providecommand \natexlab [1]{#1}%
\providecommand \enquote  [1]{``#1''}%
\providecommand \bibnamefont  [1]{#1}%
\providecommand \bibfnamefont [1]{#1}%
\providecommand \citenamefont [1]{#1}%
\providecommand \href@noop [0]{\@secondoftwo}%
\providecommand \href [0]{\begingroup \@sanitize@url \@href}%
\providecommand \@href[1]{\@@startlink{#1}\@@href}%
\providecommand \@@href[1]{\endgroup#1\@@endlink}%
\providecommand \@sanitize@url [0]{\catcode `\\12\catcode `\$12\catcode
  `\&12\catcode `\#12\catcode `\^12\catcode `\_12\catcode `\%12\relax}%
\providecommand \@@startlink[1]{}%
\providecommand \@@endlink[0]{}%
\providecommand \url  [0]{\begingroup\@sanitize@url \@url }%
\providecommand \@url [1]{\endgroup\@href {#1}{\urlprefix }}%
\providecommand \urlprefix  [0]{URL }%
\providecommand \Eprint [0]{\href }%
\providecommand \doibase [0]{http://dx.doi.org/}%
\providecommand \selectlanguage [0]{\@gobble}%
\providecommand \bibinfo  [0]{\@secondoftwo}%
\providecommand \bibfield  [0]{\@secondoftwo}%
\providecommand \translation [1]{[#1]}%
\providecommand \BibitemOpen [0]{}%
\providecommand \bibitemStop [0]{}%
\providecommand \bibitemNoStop [0]{.\EOS\space}%
\providecommand \EOS [0]{\spacefactor3000\relax}%
\providecommand \BibitemShut  [1]{\csname bibitem#1\endcsname}%
\let\auto@bib@innerbib\@empty
\bibitem [{\citenamefont {{Boltzmann}}(1872)}]{BoltzmannH}%
  \BibitemOpen
  \bibfield  {author} {\bibinfo {author} {\bibfnamefont {L.}~\bibnamefont
  {{Boltzmann}}},\ }\href@noop {} {\bibfield  {journal} {\bibinfo  {journal}
  {Sitzungsberichte Akademie der Wissenschaften}\ }\textbf {\bibinfo {volume}
  {66}},\ \bibinfo {pages} {275} (\bibinfo {year} {1872})}\BibitemShut
  {NoStop}%
\bibitem [{\citenamefont {Poincar{\'{e}}}(1890)}]{Poincare1890}%
  \BibitemOpen
  \bibfield  {author} {\bibinfo {author} {\bibfnamefont {H.}~\bibnamefont
  {Poincar{\'{e}}}},\ }\href@noop {} {\bibfield  {journal} {\bibinfo  {journal}
  {Acta Mathematica}\ }\textbf {\bibinfo {volume} {13}},\ \bibinfo {pages} {1}
  (\bibinfo {year} {1890})}\BibitemShut {NoStop}%
\bibitem [{\citenamefont {{Zermelo}}(1896)}]{Zermelo1896}%
  \BibitemOpen
  \bibfield  {author} {\bibinfo {author} {\bibfnamefont {E.}~\bibnamefont
  {{Zermelo}}},\ }\href {\doibase 10.1002/andp.18962930314} {\bibfield
  {journal} {\bibinfo  {journal} {Annalen der Physik}\ }\textbf {\bibinfo
  {volume} {293}},\ \bibinfo {pages} {485} (\bibinfo {year}
  {1896})}\BibitemShut {NoStop}%
\bibitem [{\citenamefont {{ter Haar}}(1955)}]{terHaar1955}%
  \BibitemOpen
  \bibfield  {author} {\bibinfo {author} {\bibfnamefont {D.}~\bibnamefont {{ter
  Haar}}},\ }\href {\doibase 10.1103/RevModPhys.27.289} {\bibfield  {journal}
  {\bibinfo  {journal} {Reviews of Modern Physics}\ }\textbf {\bibinfo {volume}
  {27}},\ \bibinfo {pages} {289} (\bibinfo {year} {1955})}\BibitemShut
  {NoStop}%
\bibitem [{\citenamefont {{von Neumann}}(1929)}]{Neumann1929}%
  \BibitemOpen
  \bibfield  {author} {\bibinfo {author} {\bibfnamefont {J.}~\bibnamefont {{von
  Neumann}}},\ }\href {\doibase 10.1007/BF01339852} {\bibfield  {journal}
  {\bibinfo  {journal} {Zeitschrift f\"ur Physik}\ }\textbf {\bibinfo {volume}
  {57}},\ \bibinfo {pages} {30} (\bibinfo {year} {1929})}\BibitemShut {NoStop}%
\bibitem [{\citenamefont {Gogolin}\ and\ \citenamefont
  {Eisert}(2016)}]{Gogolin2016}%
  \BibitemOpen
  \bibfield  {author} {\bibinfo {author} {\bibfnamefont {C.}~\bibnamefont
  {Gogolin}}\ and\ \bibinfo {author} {\bibfnamefont {J.}~\bibnamefont
  {Eisert}},\ }\href {\doibase 10.1088/0034-4885/79/5/056001} {\bibfield
  {journal} {\bibinfo  {journal} {Reports on Progress in Physics}\ }\textbf
  {\bibinfo {volume} {79}},\ \bibinfo {pages} {056001} (\bibinfo {year}
  {2016})}\BibitemShut {NoStop}%
\bibitem [{\citenamefont {Bocchieri}\ and\ \citenamefont
  {Loinger}(1957)}]{Bocchieri1957}%
  \BibitemOpen
  \bibfield  {author} {\bibinfo {author} {\bibfnamefont {P.}~\bibnamefont
  {Bocchieri}}\ and\ \bibinfo {author} {\bibfnamefont {A.}~\bibnamefont
  {Loinger}},\ }\href {\doibase 10.1103/PhysRev.107.337} {\bibfield  {journal}
  {\bibinfo  {journal} {Physical Review}\ }\textbf {\bibinfo {volume} {107}},\
  \bibinfo {pages} {337} (\bibinfo {year} {1957})}\BibitemShut {NoStop}%
\bibitem [{\citenamefont {Percival}(1961)}]{Percival1961}%
  \BibitemOpen
  \bibfield  {author} {\bibinfo {author} {\bibfnamefont {I.~C.}\ \bibnamefont
  {Percival}},\ }\href {\doibase 10.1063/1.1703705} {\bibfield  {journal}
  {\bibinfo  {journal} {Journal of Mathematical Physics}\ }\textbf {\bibinfo
  {volume} {2}},\ \bibinfo {pages} {235} (\bibinfo {year} {1961})}\BibitemShut
  {NoStop}%
\bibitem [{\citenamefont {Cummings}(1965)}]{Cummings1965}%
  \BibitemOpen
  \bibfield  {author} {\bibinfo {author} {\bibfnamefont {F.~W.}\ \bibnamefont
  {Cummings}},\ }\href {\doibase 10.1103/PhysRev.140.A1051} {\bibfield
  {journal} {\bibinfo  {journal} {Physical Review}\ }\textbf {\bibinfo {volume}
  {140}},\ \bibinfo {pages} {A1051} (\bibinfo {year} {1965})}\BibitemShut
  {NoStop}%
\bibitem [{\citenamefont {Eberly}\ \emph {et~al.}(1980)\citenamefont {Eberly},
  \citenamefont {Narozhny},\ and\ \citenamefont
  {Sanchez-Mondragon}}]{Eberly1980}%
  \BibitemOpen
  \bibfield  {author} {\bibinfo {author} {\bibfnamefont {J.~H.}\ \bibnamefont
  {Eberly}}, \bibinfo {author} {\bibfnamefont {N.~B.}\ \bibnamefont
  {Narozhny}}, \ and\ \bibinfo {author} {\bibfnamefont {J.~J.}\ \bibnamefont
  {Sanchez-Mondragon}},\ }\href {\doibase 10.1103/PhysRevLett.44.1323}
  {\bibfield  {journal} {\bibinfo  {journal} {Physical Review Letters}\
  }\textbf {\bibinfo {volume} {44}},\ \bibinfo {pages} {1323} (\bibinfo {year}
  {1980})}\BibitemShut {NoStop}%
\bibitem [{\citenamefont {Rempe}\ \emph {et~al.}(1987)\citenamefont {Rempe},
  \citenamefont {Walther},\ and\ \citenamefont {Klein}}]{Rempe1987}%
  \BibitemOpen
  \bibfield  {author} {\bibinfo {author} {\bibfnamefont {G.}~\bibnamefont
  {Rempe}}, \bibinfo {author} {\bibfnamefont {H.}~\bibnamefont {Walther}}, \
  and\ \bibinfo {author} {\bibfnamefont {N.}~\bibnamefont {Klein}},\ }\href
  {\doibase 10.1103/PhysRevLett.58.353} {\bibfield  {journal} {\bibinfo
  {journal} {Physical Review Letters}\ }\textbf {\bibinfo {volume} {58}},\
  \bibinfo {pages} {353} (\bibinfo {year} {1987})}\BibitemShut {NoStop}%
\bibitem [{\citenamefont {Greiner}\ \emph {et~al.}(2002)\citenamefont
  {Greiner}, \citenamefont {Mandel}, \citenamefont {H{\"{a}}nsch},\ and\
  \citenamefont {Bloch}}]{Greiner2002a}%
  \BibitemOpen
  \bibfield  {author} {\bibinfo {author} {\bibfnamefont {M.}~\bibnamefont
  {Greiner}}, \bibinfo {author} {\bibfnamefont {O.}~\bibnamefont {Mandel}},
  \bibinfo {author} {\bibfnamefont {T.~W.}\ \bibnamefont {H{\"{a}}nsch}}, \
  and\ \bibinfo {author} {\bibfnamefont {I.}~\bibnamefont {Bloch}},\ }\href
  {\doibase 10.1038/nature00968} {\bibfield  {journal} {\bibinfo  {journal}
  {Nature}\ }\textbf {\bibinfo {volume} {419}},\ \bibinfo {pages} {51}
  (\bibinfo {year} {2002})}\BibitemShut {NoStop}%
\bibitem [{\citenamefont {Will}\ \emph {et~al.}(2010)\citenamefont {Will},
  \citenamefont {Best}, \citenamefont {Schneider}, \citenamefont
  {Hackerm{\"{u}}ller}, \citenamefont {L{\"{u}}hmann},\ and\ \citenamefont
  {Bloch}}]{Will2010}%
  \BibitemOpen
  \bibfield  {author} {\bibinfo {author} {\bibfnamefont {S.}~\bibnamefont
  {Will}}, \bibinfo {author} {\bibfnamefont {T.}~\bibnamefont {Best}}, \bibinfo
  {author} {\bibfnamefont {U.}~\bibnamefont {Schneider}}, \bibinfo {author}
  {\bibfnamefont {L.}~\bibnamefont {Hackerm{\"{u}}ller}}, \bibinfo {author}
  {\bibfnamefont {D.-S.}\ \bibnamefont {L{\"{u}}hmann}}, \ and\ \bibinfo
  {author} {\bibfnamefont {I.}~\bibnamefont {Bloch}},\ }\href {\doibase
  10.1038/nature09036} {\bibfield  {journal} {\bibinfo  {journal} {Nature}\
  }\textbf {\bibinfo {volume} {465}},\ \bibinfo {pages} {197} (\bibinfo {year}
  {2010})}\BibitemShut {NoStop}%
\bibitem [{\citenamefont {Schweigler}\ \emph {et~al.}(2017)\citenamefont
  {Schweigler}, \citenamefont {Kasper}, \citenamefont {Erne}, \citenamefont
  {Mazets}, \citenamefont {Rauer}, \citenamefont {Cataldini}, \citenamefont
  {Langen}, \citenamefont {Gasenzer}, \citenamefont {Berges},\ and\
  \citenamefont {Schmiedmayer}}]{Schweigler2017}%
  \BibitemOpen
  \bibfield  {author} {\bibinfo {author} {\bibfnamefont {T.}~\bibnamefont
  {Schweigler}}, \bibinfo {author} {\bibfnamefont {V.}~\bibnamefont {Kasper}},
  \bibinfo {author} {\bibfnamefont {S.}~\bibnamefont {Erne}}, \bibinfo {author}
  {\bibfnamefont {I.~E.}\ \bibnamefont {Mazets}}, \bibinfo {author}
  {\bibfnamefont {B.}~\bibnamefont {Rauer}}, \bibinfo {author} {\bibfnamefont
  {F.}~\bibnamefont {Cataldini}}, \bibinfo {author} {\bibfnamefont
  {T.}~\bibnamefont {Langen}}, \bibinfo {author} {\bibfnamefont
  {T.}~\bibnamefont {Gasenzer}}, \bibinfo {author} {\bibfnamefont
  {J.}~\bibnamefont {Berges}}, \ and\ \bibinfo {author} {\bibfnamefont
  {J.}~\bibnamefont {Schmiedmayer}},\ }\href {\doibase 10.1038/nature22310}
  {\bibfield  {journal} {\bibinfo  {journal} {Nature}\ }\textbf {\bibinfo
  {volume} {545}},\ \bibinfo {pages} {323} (\bibinfo {year}
  {2017})}\BibitemShut {NoStop}%
\bibitem [{\citenamefont {Proukakis}\ \emph {et~al.}(2017)\citenamefont
  {Proukakis}, \citenamefont {Snoke},\ and\ \citenamefont
  {Littlewood}}]{Proukakis2017}%
  \BibitemOpen
  \bibfield  {author} {\bibinfo {author} {\bibfnamefont {N.~P.}\ \bibnamefont
  {Proukakis}}, \bibinfo {author} {\bibfnamefont {D.~W.}\ \bibnamefont
  {Snoke}}, \ and\ \bibinfo {author} {\bibfnamefont {P.~B.}\ \bibnamefont
  {Littlewood}},\ }\href@noop {} {\emph {\bibinfo {title} {Universal Themes of
  Bose-Einstein Condensation}}}\ (\bibinfo  {publisher} {Cambridge University
  Press},\ \bibinfo {address} {Cambridge},\ \bibinfo {year} {2017})\BibitemShut
  {NoStop}%
\bibitem [{\citenamefont {Tomonaga}(1950)}]{Tomonaga1950a}%
  \BibitemOpen
  \bibfield  {author} {\bibinfo {author} {\bibfnamefont {S.-I.}\ \bibnamefont
  {Tomonaga}},\ }\href {\doibase 10.1143/ptp/5.4.544} {\bibfield  {journal}
  {\bibinfo  {journal} {Progress of Theoretical Physics}\ }\textbf {\bibinfo
  {volume} {5}},\ \bibinfo {pages} {544} (\bibinfo {year} {1950})}\BibitemShut
  {NoStop}%
\bibitem [{\citenamefont {Luttinger}(1963)}]{Luttinger1963}%
  \BibitemOpen
  \bibfield  {author} {\bibinfo {author} {\bibfnamefont {J.~M.}\ \bibnamefont
  {Luttinger}},\ }\href {\doibase 10.1063/1.1704046} {\bibfield  {journal}
  {\bibinfo  {journal} {Journal of Mathematical Physics}\ }\textbf {\bibinfo
  {volume} {4}},\ \bibinfo {pages} {1154} (\bibinfo {year} {1963})}\BibitemShut
  {NoStop}%
\bibitem [{\citenamefont {Mattis}\ and\ \citenamefont
  {Lieb}(1965)}]{Mattis1965}%
  \BibitemOpen
  \bibfield  {author} {\bibinfo {author} {\bibfnamefont {D.~C.}\ \bibnamefont
  {Mattis}}\ and\ \bibinfo {author} {\bibfnamefont {E.~H.}\ \bibnamefont
  {Lieb}},\ }\href {\doibase 10.1063/1.1704281} {\bibfield  {journal} {\bibinfo
   {journal} {Journal of Mathematical Physics}\ }\textbf {\bibinfo {volume}
  {6}},\ \bibinfo {pages} {304} (\bibinfo {year} {1965})}\BibitemShut {NoStop}%
\bibitem [{\citenamefont {Giamarchi}(2004)}]{Giamarchi2004}%
  \BibitemOpen
  \bibfield  {author} {\bibinfo {author} {\bibfnamefont {T.}~\bibnamefont
  {Giamarchi}},\ }\href@noop {} {\emph {\bibinfo {title} {Quantum physics in
  one dimension}}}\ (\bibinfo  {publisher} {Clarendon Press},\ \bibinfo
  {address} {Oxford},\ \bibinfo {year} {2004})\BibitemShut {NoStop}%
\bibitem [{\citenamefont {Bistritzer}\ and\ \citenamefont
  {Altman}(2007)}]{Bistritzer2007}%
  \BibitemOpen
  \bibfield  {author} {\bibinfo {author} {\bibfnamefont {R.}~\bibnamefont
  {Bistritzer}}\ and\ \bibinfo {author} {\bibfnamefont {E.}~\bibnamefont
  {Altman}},\ }\href {\doibase 10.1073/pnas.0608910104} {\bibfield  {journal}
  {\bibinfo  {journal} {Proceedings of the National Academy of Sciences of the
  United States of America}\ }\textbf {\bibinfo {volume} {104}},\ \bibinfo
  {pages} {9955} (\bibinfo {year} {2007})}\BibitemShut {NoStop}%
\bibitem [{\citenamefont {Kitagawa}\ \emph {et~al.}(2011)\citenamefont
  {Kitagawa}, \citenamefont {Imambekov}, \citenamefont {Schmiedmayer},\ and\
  \citenamefont {Demler}}]{Kitagawa2011}%
  \BibitemOpen
  \bibfield  {author} {\bibinfo {author} {\bibfnamefont {T.}~\bibnamefont
  {Kitagawa}}, \bibinfo {author} {\bibfnamefont {A.}~\bibnamefont {Imambekov}},
  \bibinfo {author} {\bibfnamefont {J.}~\bibnamefont {Schmiedmayer}}, \ and\
  \bibinfo {author} {\bibfnamefont {E.}~\bibnamefont {Demler}},\ }\href
  {\doibase 10.1088/1367-2630/13/7/073018} {\bibfield  {journal} {\bibinfo
  {journal} {New Journal of Physics}\ }\textbf {\bibinfo {volume} {13}},\
  \bibinfo {pages} {073018} (\bibinfo {year} {2011})}\BibitemShut {NoStop}%
\bibitem [{\citenamefont {Gring}\ \emph {et~al.}(2012)\citenamefont {Gring},
  \citenamefont {Kuhnert}, \citenamefont {Langen}, \citenamefont {Kitagawa},
  \citenamefont {Rauer}, \citenamefont {Schreitl}, \citenamefont {Mazets},
  \citenamefont {Smith}, \citenamefont {Demler},\ and\ \citenamefont
  {Schmiedmayer}}]{Gring2012}%
  \BibitemOpen
  \bibfield  {author} {\bibinfo {author} {\bibfnamefont {M.}~\bibnamefont
  {Gring}}, \bibinfo {author} {\bibfnamefont {M.}~\bibnamefont {Kuhnert}},
  \bibinfo {author} {\bibfnamefont {T.}~\bibnamefont {Langen}}, \bibinfo
  {author} {\bibfnamefont {T.}~\bibnamefont {Kitagawa}}, \bibinfo {author}
  {\bibfnamefont {B.}~\bibnamefont {Rauer}}, \bibinfo {author} {\bibfnamefont
  {M.}~\bibnamefont {Schreitl}}, \bibinfo {author} {\bibfnamefont {I.~E.}\
  \bibnamefont {Mazets}}, \bibinfo {author} {\bibfnamefont {D.~A.}\
  \bibnamefont {Smith}}, \bibinfo {author} {\bibfnamefont {E.}~\bibnamefont
  {Demler}}, \ and\ \bibinfo {author} {\bibfnamefont {J.}~\bibnamefont
  {Schmiedmayer}},\ }\href {\doibase 10.1126/science.1224953} {\bibfield
  {journal} {\bibinfo  {journal} {Science}\ }\textbf {\bibinfo {volume}
  {337}},\ \bibinfo {pages} {1318} (\bibinfo {year} {2012})}\BibitemShut
  {NoStop}%
\bibitem [{\citenamefont {Langen}\ \emph {et~al.}(2013)\citenamefont {Langen},
  \citenamefont {Geiger}, \citenamefont {Kuhnert}, \citenamefont {Rauer},\ and\
  \citenamefont {Schmiedmayer}}]{Langen2013}%
  \BibitemOpen
  \bibfield  {author} {\bibinfo {author} {\bibfnamefont {T.}~\bibnamefont
  {Langen}}, \bibinfo {author} {\bibfnamefont {R.}~\bibnamefont {Geiger}},
  \bibinfo {author} {\bibfnamefont {M.}~\bibnamefont {Kuhnert}}, \bibinfo
  {author} {\bibfnamefont {B.}~\bibnamefont {Rauer}}, \ and\ \bibinfo {author}
  {\bibfnamefont {J.}~\bibnamefont {Schmiedmayer}},\ }\href {\doibase
  10.1038/nphys2739} {\bibfield  {journal} {\bibinfo  {journal} {Nature
  Physics}\ }\textbf {\bibinfo {volume} {9}},\ \bibinfo {pages} {640} (\bibinfo
  {year} {2013})}\BibitemShut {NoStop}%
\bibitem [{\citenamefont {Langen}\ \emph {et~al.}(2015)\citenamefont {Langen},
  \citenamefont {Erne}, \citenamefont {Geiger}, \citenamefont {Rauer},
  \citenamefont {Schweigler}, \citenamefont {Kuhnert}, \citenamefont
  {Rohringer}, \citenamefont {Mazets}, \citenamefont {Gasenzer},\ and\
  \citenamefont {Schmiedmayer}}]{Langen2015a}%
  \BibitemOpen
  \bibfield  {author} {\bibinfo {author} {\bibfnamefont {T.}~\bibnamefont
  {Langen}}, \bibinfo {author} {\bibfnamefont {S.}~\bibnamefont {Erne}},
  \bibinfo {author} {\bibfnamefont {R.}~\bibnamefont {Geiger}}, \bibinfo
  {author} {\bibfnamefont {B.}~\bibnamefont {Rauer}}, \bibinfo {author}
  {\bibfnamefont {T.}~\bibnamefont {Schweigler}}, \bibinfo {author}
  {\bibfnamefont {M.}~\bibnamefont {Kuhnert}}, \bibinfo {author} {\bibfnamefont
  {W.}~\bibnamefont {Rohringer}}, \bibinfo {author} {\bibfnamefont {I.~E.}\
  \bibnamefont {Mazets}}, \bibinfo {author} {\bibfnamefont {T.}~\bibnamefont
  {Gasenzer}}, \ and\ \bibinfo {author} {\bibfnamefont {J.}~\bibnamefont
  {Schmiedmayer}},\ }\href {\doibase 10.1126/science.1257026} {\bibfield
  {journal} {\bibinfo  {journal} {Science}\ }\textbf {\bibinfo {volume}
  {348}},\ \bibinfo {pages} {207} (\bibinfo {year} {2015})}\BibitemShut
  {NoStop}%
\bibitem [{\citenamefont {Petrov}\ \emph {et~al.}(2000)\citenamefont {Petrov},
  \citenamefont {Shlyapnikov},\ and\ \citenamefont {Walraven}}]{Petrov2000}%
  \BibitemOpen
  \bibfield  {author} {\bibinfo {author} {\bibfnamefont {D.~S.}\ \bibnamefont
  {Petrov}}, \bibinfo {author} {\bibfnamefont {G.~V.}\ \bibnamefont
  {Shlyapnikov}}, \ and\ \bibinfo {author} {\bibfnamefont {J.~T.~M.}\
  \bibnamefont {Walraven}},\ }\href {\doibase 10.1103/PhysRevLett.85.3745}
  {\bibfield  {journal} {\bibinfo  {journal} {Physical Review Letters}\
  }\textbf {\bibinfo {volume} {85}},\ \bibinfo {pages} {3745} (\bibinfo {year}
  {2000})}\BibitemShut {NoStop}%
\bibitem [{\citenamefont {Geiger}\ \emph {et~al.}(2014)\citenamefont {Geiger},
  \citenamefont {Langen}, \citenamefont {Mazets},\ and\ \citenamefont
  {Schmiedmayer}}]{Geiger2014}%
  \BibitemOpen
  \bibfield  {author} {\bibinfo {author} {\bibfnamefont {R.}~\bibnamefont
  {Geiger}}, \bibinfo {author} {\bibfnamefont {T.}~\bibnamefont {Langen}},
  \bibinfo {author} {\bibfnamefont {I.~E.}\ \bibnamefont {Mazets}}, \ and\
  \bibinfo {author} {\bibfnamefont {J.}~\bibnamefont {Schmiedmayer}},\ }\href
  {\doibase 10.1088/1367-2630/16/5/053034} {\bibfield  {journal} {\bibinfo
  {journal} {New Journal of Physics}\ }\textbf {\bibinfo {volume} {16}},\
  \bibinfo {pages} {053034} (\bibinfo {year} {2014})}\BibitemShut {NoStop}%
\bibitem [{\citenamefont {Schumm}\ \emph {et~al.}(2005)\citenamefont {Schumm},
  \citenamefont {Hofferberth}, \citenamefont {Andersson}, \citenamefont
  {Wildermuth}, \citenamefont {Groth}, \citenamefont {Bar-Joseph},
  \citenamefont {Schmiedmayer},\ and\ \citenamefont
  {Kr{\"{u}}ger}}]{Schumm2005}%
  \BibitemOpen
  \bibfield  {author} {\bibinfo {author} {\bibfnamefont {T.}~\bibnamefont
  {Schumm}}, \bibinfo {author} {\bibfnamefont {S.}~\bibnamefont {Hofferberth}},
  \bibinfo {author} {\bibfnamefont {L.~M.}\ \bibnamefont {Andersson}}, \bibinfo
  {author} {\bibfnamefont {S.}~\bibnamefont {Wildermuth}}, \bibinfo {author}
  {\bibfnamefont {S.}~\bibnamefont {Groth}}, \bibinfo {author} {\bibfnamefont
  {I.}~\bibnamefont {Bar-Joseph}}, \bibinfo {author} {\bibfnamefont
  {J.}~\bibnamefont {Schmiedmayer}}, \ and\ \bibinfo {author} {\bibfnamefont
  {P.}~\bibnamefont {Kr{\"{u}}ger}},\ }\href {\doibase 10.1038/nphys125}
  {\bibfield  {journal} {\bibinfo  {journal} {Nature Physics}\ }\textbf
  {\bibinfo {volume} {1}},\ \bibinfo {pages} {57} (\bibinfo {year}
  {2005})}\BibitemShut {NoStop}%
\bibitem [{\citenamefont {Betz}\ \emph {et~al.}(2011)\citenamefont {Betz},
  \citenamefont {Manz}, \citenamefont {B{\"{u}}cker}, \citenamefont {Berrada},
  \citenamefont {Koller}, \citenamefont {Kazakov}, \citenamefont {Mazets},
  \citenamefont {Stimming}, \citenamefont {Perrin}, \citenamefont {Schumm},\
  and\ \citenamefont {Schmiedmayer}}]{Betz2011}%
  \BibitemOpen
  \bibfield  {author} {\bibinfo {author} {\bibfnamefont {T.}~\bibnamefont
  {Betz}}, \bibinfo {author} {\bibfnamefont {S.}~\bibnamefont {Manz}}, \bibinfo
  {author} {\bibfnamefont {R.}~\bibnamefont {B{\"{u}}cker}}, \bibinfo {author}
  {\bibfnamefont {T.}~\bibnamefont {Berrada}}, \bibinfo {author} {\bibfnamefont
  {C.}~\bibnamefont {Koller}}, \bibinfo {author} {\bibfnamefont
  {G.}~\bibnamefont {Kazakov}}, \bibinfo {author} {\bibfnamefont {I.~E.}\
  \bibnamefont {Mazets}}, \bibinfo {author} {\bibfnamefont {H.-P.}\
  \bibnamefont {Stimming}}, \bibinfo {author} {\bibfnamefont {A.}~\bibnamefont
  {Perrin}}, \bibinfo {author} {\bibfnamefont {T.}~\bibnamefont {Schumm}}, \
  and\ \bibinfo {author} {\bibfnamefont {J.}~\bibnamefont {Schmiedmayer}},\
  }\href {\doibase 10.1103/PhysRevLett.106.020407} {\bibfield  {journal}
  {\bibinfo  {journal} {Physical Review Letters}\ }\textbf {\bibinfo {volume}
  {106}},\ \bibinfo {pages} {020407} (\bibinfo {year} {2011})}\BibitemShut
  {NoStop}%
\bibitem [{\citenamefont {Hofferberth}\ \emph {et~al.}(2008)\citenamefont
  {Hofferberth}, \citenamefont {Lesanovsky}, \citenamefont {Schumm},
  \citenamefont {Imambekov}, \citenamefont {Gritsev}, \citenamefont {Demler},\
  and\ \citenamefont {Schmiedmayer}}]{Hofferberth2008}%
  \BibitemOpen
  \bibfield  {author} {\bibinfo {author} {\bibfnamefont {S.}~\bibnamefont
  {Hofferberth}}, \bibinfo {author} {\bibfnamefont {I.}~\bibnamefont
  {Lesanovsky}}, \bibinfo {author} {\bibfnamefont {T.}~\bibnamefont {Schumm}},
  \bibinfo {author} {\bibfnamefont {A.}~\bibnamefont {Imambekov}}, \bibinfo
  {author} {\bibfnamefont {V.}~\bibnamefont {Gritsev}}, \bibinfo {author}
  {\bibfnamefont {E.}~\bibnamefont {Demler}}, \ and\ \bibinfo {author}
  {\bibfnamefont {J.}~\bibnamefont {Schmiedmayer}},\ }\href {\doibase
  10.1038/nphys941} {\bibfield  {journal} {\bibinfo  {journal} {Nature
  Physics}\ }\textbf {\bibinfo {volume} {4}},\ \bibinfo {pages} {489} (\bibinfo
  {year} {2008})}\BibitemShut {NoStop}%
\bibitem [{\citenamefont {Cardy}(2014)}]{Cardy2014}%
  \BibitemOpen
  \bibfield  {author} {\bibinfo {author} {\bibfnamefont {J.}~\bibnamefont
  {Cardy}},\ }\href {\doibase 10.1103/PhysRevLett.112.220401} {\bibfield
  {journal} {\bibinfo  {journal} {Physical Review Letters}\ }\textbf {\bibinfo
  {volume} {112}},\ \bibinfo {pages} {220401} (\bibinfo {year}
  {2014})}\BibitemShut {NoStop}%
\bibitem [{\citenamefont {Folman}\ \emph {et~al.}(2000)\citenamefont {Folman},
  \citenamefont {Kr{\"{u}}ger}, \citenamefont {Cassettari}, \citenamefont
  {Hessmo}, \citenamefont {Maier},\ and\ \citenamefont
  {Schmiedmayer}}]{Folman2000}%
  \BibitemOpen
  \bibfield  {author} {\bibinfo {author} {\bibfnamefont {R.}~\bibnamefont
  {Folman}}, \bibinfo {author} {\bibfnamefont {P.}~\bibnamefont
  {Kr{\"{u}}ger}}, \bibinfo {author} {\bibfnamefont {D.}~\bibnamefont
  {Cassettari}}, \bibinfo {author} {\bibfnamefont {B.}~\bibnamefont {Hessmo}},
  \bibinfo {author} {\bibfnamefont {T.}~\bibnamefont {Maier}}, \ and\ \bibinfo
  {author} {\bibfnamefont {J.}~\bibnamefont {Schmiedmayer}},\ }\href {\doibase
  10.1103/PhysRevLett.84.4749} {\bibfield  {journal} {\bibinfo  {journal}
  {Physical Review Letters}\ }\textbf {\bibinfo {volume} {84}},\ \bibinfo
  {pages} {4749} (\bibinfo {year} {2000})}\BibitemShut {NoStop}%
\bibitem [{\citenamefont {Reichel}\ and\ \citenamefont
  {Vuleti\'c}(2011)}]{Reichel2011}%
  \BibitemOpen
  \bibfield  {author} {\bibinfo {author} {\bibfnamefont {J.}~\bibnamefont
  {Reichel}}\ and\ \bibinfo {author} {\bibfnamefont {V.}~\bibnamefont
  {Vuleti\'c}},\ }\href@noop {} {\emph {\bibinfo {title} {Atom Chips}}}\
  (\bibinfo  {publisher} {Wiley-VCH},\ \bibinfo {address} {Weinheim, Germany},\
  \bibinfo {year} {2011})\BibitemShut {NoStop}%
\bibitem [{\citenamefont {Hofferberth}\ \emph {et~al.}(2006)\citenamefont
  {Hofferberth}, \citenamefont {Lesanovsky}, \citenamefont {Fischer},
  \citenamefont {Verdu},\ and\ \citenamefont {Schmiedmayer}}]{Hofferberth2006}%
  \BibitemOpen
  \bibfield  {author} {\bibinfo {author} {\bibfnamefont {S.}~\bibnamefont
  {Hofferberth}}, \bibinfo {author} {\bibfnamefont {I.}~\bibnamefont
  {Lesanovsky}}, \bibinfo {author} {\bibfnamefont {B.}~\bibnamefont {Fischer}},
  \bibinfo {author} {\bibfnamefont {J.}~\bibnamefont {Verdu}}, \ and\ \bibinfo
  {author} {\bibfnamefont {J.}~\bibnamefont {Schmiedmayer}},\ }\href {\doibase
  10.1038/nphys420} {\bibfield  {journal} {\bibinfo  {journal} {Nature
  Physics}\ }\textbf {\bibinfo {volume} {2}},\ \bibinfo {pages} {710} (\bibinfo
  {year} {2006})}\BibitemShut {NoStop}%
\bibitem [{\citenamefont {Lesanovsky}\ \emph {et~al.}(2006)\citenamefont
  {Lesanovsky}, \citenamefont {Hofferberth}, \citenamefont {Schmiedmayer},\
  and\ \citenamefont {Schmelcher}}]{Lesanovsky2006a}%
  \BibitemOpen
  \bibfield  {author} {\bibinfo {author} {\bibfnamefont {I.}~\bibnamefont
  {Lesanovsky}}, \bibinfo {author} {\bibfnamefont {S.}~\bibnamefont
  {Hofferberth}}, \bibinfo {author} {\bibfnamefont {J.}~\bibnamefont
  {Schmiedmayer}}, \ and\ \bibinfo {author} {\bibfnamefont {P.}~\bibnamefont
  {Schmelcher}},\ }\href {\doibase 10.1103/PhysRevA.74.033619} {\bibfield
  {journal} {\bibinfo  {journal} {Physical Review A}\ }\textbf {\bibinfo
  {volume} {74}},\ \bibinfo {pages} {033619} (\bibinfo {year}
  {2006})}\BibitemShut {NoStop}%
\bibitem [{\citenamefont {Smith}\ \emph {et~al.}(2013)\citenamefont {Smith},
  \citenamefont {Gring}, \citenamefont {Langen}, \citenamefont {Kuhnert},
  \citenamefont {Rauer}, \citenamefont {Geiger}, \citenamefont {Kitagawa},
  \citenamefont {Mazets}, \citenamefont {Demler},\ and\ \citenamefont
  {Schmiedmayer}}]{AduSmith2013a}%
  \BibitemOpen
  \bibfield  {author} {\bibinfo {author} {\bibfnamefont {D.~A.}\ \bibnamefont
  {Smith}}, \bibinfo {author} {\bibfnamefont {M.}~\bibnamefont {Gring}},
  \bibinfo {author} {\bibfnamefont {T.}~\bibnamefont {Langen}}, \bibinfo
  {author} {\bibfnamefont {M.}~\bibnamefont {Kuhnert}}, \bibinfo {author}
  {\bibfnamefont {B.}~\bibnamefont {Rauer}}, \bibinfo {author} {\bibfnamefont
  {R.}~\bibnamefont {Geiger}}, \bibinfo {author} {\bibfnamefont
  {T.}~\bibnamefont {Kitagawa}}, \bibinfo {author} {\bibfnamefont {I.~E.}\
  \bibnamefont {Mazets}}, \bibinfo {author} {\bibfnamefont {E.}~\bibnamefont
  {Demler}}, \ and\ \bibinfo {author} {\bibfnamefont {J.}~\bibnamefont
  {Schmiedmayer}},\ }\href {\doibase 10.1088/1367-2630/15/7/075011} {\bibfield
  {journal} {\bibinfo  {journal} {New Journal of Physics}\ }\textbf {\bibinfo
  {volume} {15}},\ \bibinfo {pages} {075011} (\bibinfo {year}
  {2013})}\BibitemShut {NoStop}%
\bibitem [{\citenamefont {Manz}\ \emph {et~al.}(2010)\citenamefont {Manz},
  \citenamefont {B{\"{u}}cker}, \citenamefont {Betz}, \citenamefont {Koller},
  \citenamefont {Hofferberth}, \citenamefont {Mazets}, \citenamefont
  {Imambekov}, \citenamefont {Demler}, \citenamefont {Perrin}, \citenamefont
  {Schmiedmayer},\ and\ \citenamefont {Schumm}}]{Manz2010}%
  \BibitemOpen
  \bibfield  {author} {\bibinfo {author} {\bibfnamefont {S.}~\bibnamefont
  {Manz}}, \bibinfo {author} {\bibfnamefont {R.}~\bibnamefont {B{\"{u}}cker}},
  \bibinfo {author} {\bibfnamefont {T.}~\bibnamefont {Betz}}, \bibinfo {author}
  {\bibfnamefont {C.}~\bibnamefont {Koller}}, \bibinfo {author} {\bibfnamefont
  {S.}~\bibnamefont {Hofferberth}}, \bibinfo {author} {\bibfnamefont {I.~E.}\
  \bibnamefont {Mazets}}, \bibinfo {author} {\bibfnamefont {A.}~\bibnamefont
  {Imambekov}}, \bibinfo {author} {\bibfnamefont {E.}~\bibnamefont {Demler}},
  \bibinfo {author} {\bibfnamefont {A.}~\bibnamefont {Perrin}}, \bibinfo
  {author} {\bibfnamefont {J.}~\bibnamefont {Schmiedmayer}}, \ and\ \bibinfo
  {author} {\bibfnamefont {T.}~\bibnamefont {Schumm}},\ }\href {\doibase
  10.1103/PhysRevA.81.031610} {\bibfield  {journal} {\bibinfo  {journal}
  {Physical Review A}\ }\textbf {\bibinfo {volume} {81}},\ \bibinfo {pages}
  {031610} (\bibinfo {year} {2010})}\BibitemShut {NoStop}%
\bibitem [{\citenamefont {Imambekov}\ \emph {et~al.}(2009)\citenamefont
  {Imambekov}, \citenamefont {Mazets}, \citenamefont {Petrov}, \citenamefont
  {Gritsev}, \citenamefont {Manz}, \citenamefont {Hofferberth}, \citenamefont
  {Schumm}, \citenamefont {Demler},\ and\ \citenamefont
  {Schmiedmayer}}]{Imambekov2009}%
  \BibitemOpen
  \bibfield  {author} {\bibinfo {author} {\bibfnamefont {A.}~\bibnamefont
  {Imambekov}}, \bibinfo {author} {\bibfnamefont {I.~E.}\ \bibnamefont
  {Mazets}}, \bibinfo {author} {\bibfnamefont {D.~S.}\ \bibnamefont {Petrov}},
  \bibinfo {author} {\bibfnamefont {V.}~\bibnamefont {Gritsev}}, \bibinfo
  {author} {\bibfnamefont {S.}~\bibnamefont {Manz}}, \bibinfo {author}
  {\bibfnamefont {S.}~\bibnamefont {Hofferberth}}, \bibinfo {author}
  {\bibfnamefont {T.}~\bibnamefont {Schumm}}, \bibinfo {author} {\bibfnamefont
  {E.}~\bibnamefont {Demler}}, \ and\ \bibinfo {author} {\bibfnamefont
  {J.}~\bibnamefont {Schmiedmayer}},\ }\href {\doibase
  10.1103/PhysRevA.80.033604} {\bibfield  {journal} {\bibinfo  {journal}
  {Physical Review A}\ }\textbf {\bibinfo {volume} {80}},\ \bibinfo {pages}
  {033604} (\bibinfo {year} {2009})}\BibitemShut {NoStop}%
\bibitem [{\citenamefont {Polkovnikov}\ \emph {et~al.}(2006)\citenamefont
  {Polkovnikov}, \citenamefont {Altman},\ and\ \citenamefont
  {Demler}}]{Polkovnikov2006}%
  \BibitemOpen
  \bibfield  {author} {\bibinfo {author} {\bibfnamefont {A.}~\bibnamefont
  {Polkovnikov}}, \bibinfo {author} {\bibfnamefont {E.}~\bibnamefont {Altman}},
  \ and\ \bibinfo {author} {\bibfnamefont {E.}~\bibnamefont {Demler}},\ }\href
  {\doibase 10.1073/pnas.0510276103} {\bibfield  {journal} {\bibinfo  {journal}
  {Proceedings of the National Academy of Sciences}\ }\textbf {\bibinfo
  {volume} {103}},\ \bibinfo {pages} {6125} (\bibinfo {year}
  {2006})}\BibitemShut {NoStop}%
\bibitem [{\citenamefont {Gritsev}\ \emph {et~al.}(2006)\citenamefont
  {Gritsev}, \citenamefont {Altman}, \citenamefont {Demler},\ and\
  \citenamefont {Polkovnikov}}]{gritsev2006}%
  \BibitemOpen
  \bibfield  {author} {\bibinfo {author} {\bibfnamefont {V.}~\bibnamefont
  {Gritsev}}, \bibinfo {author} {\bibfnamefont {E.}~\bibnamefont {Altman}},
  \bibinfo {author} {\bibfnamefont {E.}~\bibnamefont {Demler}}, \ and\ \bibinfo
  {author} {\bibfnamefont {A.}~\bibnamefont {Polkovnikov}},\ }\href {\doibase
  doi:10.1038/nphys410} {\bibfield  {journal} {\bibinfo  {journal} {Nature
  Physics}\ }\textbf {\bibinfo {volume} {2}},\ \bibinfo {pages} {705} (\bibinfo
  {year} {2006})}\BibitemShut {NoStop}%
\bibitem [{\citenamefont {Stimming}\ \emph {et~al.}(2010)\citenamefont
  {Stimming}, \citenamefont {Mauser}, \citenamefont {Schmiedmayer},\ and\
  \citenamefont {Mazets}}]{Stimming2010}%
  \BibitemOpen
  \bibfield  {author} {\bibinfo {author} {\bibfnamefont {H.-P.}\ \bibnamefont
  {Stimming}}, \bibinfo {author} {\bibfnamefont {N.~J.}\ \bibnamefont
  {Mauser}}, \bibinfo {author} {\bibfnamefont {J.}~\bibnamefont
  {Schmiedmayer}}, \ and\ \bibinfo {author} {\bibfnamefont {I.~E.}\
  \bibnamefont {Mazets}},\ }\href {\doibase 10.1103/PhysRevLett.105.015301}
  {\bibfield  {journal} {\bibinfo  {journal} {Physical Review Letters}\
  }\textbf {\bibinfo {volume} {105}},\ \bibinfo {pages} {015301} (\bibinfo
  {year} {2010})}\BibitemShut {NoStop}%
\bibitem [{\citenamefont {Olshanii}(1998)}]{Olshanii1998}%
  \BibitemOpen
  \bibfield  {author} {\bibinfo {author} {\bibfnamefont {M.}~\bibnamefont
  {Olshanii}},\ }\href {\doibase 10.1103/PhysRevLett.81.938} {\bibfield
  {journal} {\bibinfo  {journal} {Physical Review Letters}\ }\textbf {\bibinfo
  {volume} {81}},\ \bibinfo {pages} {938} (\bibinfo {year} {1998})}\BibitemShut
  {NoStop}%
\bibitem [{\citenamefont {Salasnich}\ \emph {et~al.}(2002)\citenamefont
  {Salasnich}, \citenamefont {Parola},\ and\ \citenamefont
  {Reatto}}]{Salasnich2002}%
  \BibitemOpen
  \bibfield  {author} {\bibinfo {author} {\bibfnamefont {L.}~\bibnamefont
  {Salasnich}}, \bibinfo {author} {\bibfnamefont {A.}~\bibnamefont {Parola}}, \
  and\ \bibinfo {author} {\bibfnamefont {L.}~\bibnamefont {Reatto}},\ }\href
  {\doibase 10.1103/PhysRevA.65.043614} {\bibfield  {journal} {\bibinfo
  {journal} {Physical Review A}\ }\textbf {\bibinfo {volume} {65}},\ \bibinfo
  {pages} {043614} (\bibinfo {year} {2002})}\BibitemShut {NoStop}%
\bibitem [{\citenamefont {Mora}\ and\ \citenamefont {Castin}(2003)}]{Mora2003}%
  \BibitemOpen
  \bibfield  {author} {\bibinfo {author} {\bibfnamefont {C.}~\bibnamefont
  {Mora}}\ and\ \bibinfo {author} {\bibfnamefont {Y.}~\bibnamefont {Castin}},\
  }\href {\doibase 10.1103/PhysRevA.67.053615} {\bibfield  {journal} {\bibinfo
  {journal} {Physical Review A}\ }\textbf {\bibinfo {volume} {67}},\ \bibinfo
  {pages} {053615} (\bibinfo {year} {2003})}\BibitemShut {NoStop}%
\bibitem [{\citenamefont {Stoof}(1999)}]{Stoof1999}%
  \BibitemOpen
  \bibfield  {author} {\bibinfo {author} {\bibfnamefont {H.~T.~C.}\
  \bibnamefont {Stoof}},\ }\href {\doibase 10.1023/A:1021897703053} {\bibfield
  {journal} {\bibinfo  {journal} {Journal of Low Temperature Physics}\ }\textbf
  {\bibinfo {volume} {114}},\ \bibinfo {pages} {11} (\bibinfo {year}
  {1999})}\BibitemShut {NoStop}%
\bibitem [{\citenamefont {Davis}\ \emph {et~al.}(2001)\citenamefont {Davis},
  \citenamefont {Morgan},\ and\ \citenamefont {Burnett}}]{Davis2001}%
  \BibitemOpen
  \bibfield  {author} {\bibinfo {author} {\bibfnamefont {M.~J.}\ \bibnamefont
  {Davis}}, \bibinfo {author} {\bibfnamefont {S.~A.}\ \bibnamefont {Morgan}}, \
  and\ \bibinfo {author} {\bibfnamefont {K.}~\bibnamefont {Burnett}},\ }\href
  {\doibase 10.1103/PhysRevLett.87.160402} {\bibfield  {journal} {\bibinfo
  {journal} {Physical Review Letters}\ }\textbf {\bibinfo {volume} {87}},\
  \bibinfo {pages} {160402} (\bibinfo {year} {2001})}\BibitemShut {NoStop}%
\bibitem [{\citenamefont {Gardiner}\ \emph {et~al.}(2002)\citenamefont
  {Gardiner}, \citenamefont {Anglin},\ and\ \citenamefont
  {Fudge}}]{Gardiner2002}%
  \BibitemOpen
  \bibfield  {author} {\bibinfo {author} {\bibfnamefont {C.~W.}\ \bibnamefont
  {Gardiner}}, \bibinfo {author} {\bibfnamefont {J.~R.}\ \bibnamefont
  {Anglin}}, \ and\ \bibinfo {author} {\bibfnamefont {T.~I.~A.}\ \bibnamefont
  {Fudge}},\ }\href {\doibase 10.1088/0953-4075/35/6/310} {\bibfield  {journal}
  {\bibinfo  {journal} {Journal of Physics B: Atomic, Molecular and Optical
  Physics}\ }\textbf {\bibinfo {volume} {35}},\ \bibinfo {pages} {1555}
  (\bibinfo {year} {2002})}\BibitemShut {NoStop}%
\end{thebibliography}%
